\PassOptionsToPackage{unicode}{hyperref}
\PassOptionsToPackage{hyphens}{url}
\documentclass[
  11pt,
  a4paper,
]{article}
\usepackage{amsmath,amssymb}
\usepackage{iftex}
\ifPDFTeX
  \usepackage[T1]{fontenc}
  \usepackage[utf8]{inputenc}
  \usepackage{textcomp} 
\else 
  \usepackage{unicode-math} 
  \defaultfontfeatures{Scale=MatchLowercase}
  \defaultfontfeatures[\rmfamily]{Ligatures=TeX,Scale=1}
\fi
\usepackage{lmodern}
\ifPDFTeX\else
\fi
\IfFileExists{upquote.sty}{\usepackage{upquote}}{}
\IfFileExists{microtype.sty}{
  \usepackage[]{microtype}
  \UseMicrotypeSet[protrusion]{basicmath} 
}{}
\makeatletter
\@ifundefined{KOMAClassName}{
  \IfFileExists{parskip.sty}{%
    \usepackage{parskip}
  }{
    \setlength{\parindent}{0pt}
    \setlength{\parskip}{6pt plus 2pt minus 1pt}}
}{
  \KOMAoptions{parskip=half}}
\makeatother
\usepackage{xcolor}
\usepackage{longtable,booktabs,array}
\usepackage{calc} 
\usepackage{etoolbox}
\makeatletter
\patchcmd\longtable{\par}{\if@noskipsec\mbox{}\fi\par}{}{}
\makeatother
\IfFileExists{footnotehyper.sty}{\usepackage{footnotehyper}}{\usepackage{footnote}}
\makesavenoteenv{longtable}
\providecommand{\tightlist}{%
  \setlength{\itemsep}{0pt}\setlength{\parskip}{0pt}}
\usepackage{amsmath,amssymb,mathtools,bm,mathrsfs}
\usepackage{graphicx}
\usepackage{mathpazo}
\usepackage{microtype}
\usepackage{xurl}
\usepackage{setspace}
\usepackage{enumitem}
\usepackage{booktabs,longtable,array}
\usepackage{etoolbox}
\usepackage{geometry}
\usepackage{hyperref}
\setlist{nosep}
\AtBeginEnvironment{longtable}{\scriptsize}
\hypersetup{colorlinks=true,linkcolor=black,citecolor=black,urlcolor=blue}
\ifLuaTeX
  \usepackage{selnolig}  
\fi
\IfFileExists{bookmark.sty}{\usepackage{bookmark}}{\usepackage{hyperref}}
\IfFileExists{xurl.sty}{\usepackage{xurl}}{} 
\hypersetup{
  pdftitle={Uniform Inference and Certified Capacity at a Reflexive Stability Boundary},
  pdfauthor={Alejandro Rodriguez Dominguez},
  pdfsubject={Public preprint, 2 September 2026},
  hidelinks,
  pdfcreator={LaTeX via pandoc}}

\title{Uniform Inference and Certified Capacity at a Reflexive Stability
Boundary}
\author{Alejandro Rodr\'iguez Dom\'inguez\\
\small Quantitative Analysis and Artificial Intelligence Department, Miralta Finance Bank S.A.,\\
\small Madrid, Spain\\
\small Department of Computer Science, University of Reading, Reading, United Kingdom\\
\small Department of Data and AI, Albert School,\\
\small Paris, France\\
\small Corresponding author: \texttt{arodriguez@miraltabank.com}}
\date{2 September 2026}

\begin{document}
\maketitle

\hypertarget{abstract}{%
\subsection*{Abstract}\label{abstract}}
\addcontentsline{toc}{subsection}{Abstract}

This paper develops uniform inference and certified capacity decisions for an
estimated financial stability boundary. Conditional risk, temporary
cross-impact, and effective risk-bearing capacity are jointly estimated from
dependent observations. Conventional pointwise inference is reliable at a
separated simple spectral root but can fail near semisimple or defective
collisions. Projecting a valid joint confidence region for the underlying
inputs avoids this local approximation and yields a three-way regime decision
with abstention and a one-sided capacity bound. A verified two-dimensional
implementation keeps numerical error from creating a resolved sign.
Structural simulations recover the predicted tradeoff between coverage and
resolution, while observed-risk stresses distinguish statistical abstention
from an insufficient computational budget. The financial conclusions remain
conditional on the identification of cross-impact, the normalization of
capacity, and stability of the inputs over the action horizon.

\begin{center}\rule{0.5\linewidth}{0.5pt}\end{center}

\textbf{Keywords:} cross-impact; spectral inference; nonregularity;
instrumental variables; verified computation; capacity control\\
\textbf{JEL classification:} C13; C15; C32; G11; G12

\hypertarget{introduction}{%
\section*{1. Introduction}\label{introduction}}
\addcontentsline{toc}{section}{1. Introduction}

Persistent positions can feed back into expected returns through temporary
cross-impact. Empirical work documents multivariate propagation of order flow
and the covariance transmitted through trading, while execution models relate
impact, decay, and trading costs to restrictions that exclude dynamic
arbitrage (Benzaquen et al., 2017; Gatheral, 2010; Schneider and Lillo, 2019).
These mechanisms can produce an equilibrium that changes regime at a spectral
boundary. In applications, however, the conditional-risk,
cross-impact, and capacity inputs that locate the boundary must be estimated
from dependent market observations. Treating the resulting operator as known
can turn sampling error into a false regime declaration or an excessive
capacity decision.

This paper forms the inference and risk-control component of a
representation-based research programme. Rodríguez Domínguez (2026a) studies
portfolio policy when the common-driver geometry changes; Rodríguez Domínguez
(2026b) develops the equilibrium interaction among shared representations,
crowding, cross-impact, and capacity and derives the stability boundary used
here; and Rodríguez Domínguez (2026c) studies how costly replacement of a
maintained representation can appear in executed order-flow persistence. The
present paper adds joint inference on the boundary inputs, uniform regime
classification, and a capacity decision that remains valid when the active
spectral root is nonregular.

The inferential results do not depend on the existence, uniqueness, or
selection claims of the antecedent equilibrium model. Their formal input is
the estimated feedback operator and a valid joint confidence region for its
primitive components. The antecedent supplies the financial interpretation of
that operator, but any financial feedback model producing the same boundary
can use the results below. The normalization of capacity remains specific to
the application and must be defended independently.

The statistical difficulty is not confined to conventional standard errors.
Away from eigenvalue collisions, the active spectral root is locally smooth
and ordinary delta-method inference is informative. Near a collision, its
identity and sensitivity can change discontinuously; at a defective collision,
the boundary margin may converge more slowly than the underlying parameter
estimates. These failures are instances of the nonregular behavior of spectral
functionals studied by Kato (1995), Burke and Overton (2001), and Fang and
Santos (2019). A diagnostic that first selects a smooth formula and then
reports a confidence interval is therefore not reliable uniformly over the
relevant parameter class.

The paper addresses this problem by estimating the market inputs jointly,
including their long-run cross-covariances, and projecting a confidence region
for those inputs through the spectral boundary. Projection avoids local
differentiation of the boundary functional and supports a three-way decision:
subcritical, supercritical, or unresolved. The same confidence set supplies a
one-sided bound for a subsequent capacity choice. Numerical error is included
in the reported bounds, so it can widen or leave unresolved a statistical
decision but cannot create evidence about the economic regime.

The paper contributes a joint inferential and decision framework for an
estimated financial stability boundary. It embeds conditional risk,
cross-impact, and capacity in one dependent-data experiment that preserves
their long-run cross-covariances. Semisimple and defective sequences expose
the failure of pointwise spectral inference, while projection of the
primitive confidence region produces a classifier that can remain unresolved
when the data do not determine the regime. The same confidence object supplies
a one-sided capacity decision, ensuring that classification and action use a
common assessment of uncertainty. The computational study separates
statistical nonresolution from incomplete numerical certification and
evaluates the resulting tradeoff between protected capacity and operating
risk. It does not assert a universal impact model or a welfare-optimal policy.

The remainder of the paper has four sections. Section 2 positions the analysis
in the related literature. Section 3 presents the model, the regular and
nonregular inferential results, projected classification, and capacity bounds.
Section 4 reports the computational design, certification checks, evidence,
and limitations. Section 5 concludes.

\hypertarget{related-literature}{%
\section*{2. Related literature and positioning}\label{related-literature}}
\addcontentsline{toc}{section}{2. Related literature and positioning}

Empirical cross-impact studies document multivariate propagation of order flow
and the covariance transmitted through trading (Benzaquen et al., 2017).
Execution and propagator models restrict the joint behavior of impact, decay,
and trading cost through the exclusion of dynamic arbitrage (Gatheral, 2010;
Schneider and Lillo, 2019). Portfolio-execution analyses further show how
cross-impact changes the allocation and timing of trades (Mastromatteo et al.,
2017; Tomas et al., 2022). This literature disciplines the economic meaning
and admissible structure of the impact operator. The present problem differs
because the decision boundary also depends on conditional risk and an
application-specific capacity normalization, all estimated from the same
dependent observations.

The distinction between nominal and robust stability has a long history in
control theory. Stability radii quantify the smallest admissible perturbation
that can move a system across its stability boundary (Hinrichsen and
Pritchard, 1986). Related work on financial model risk studies how ambiguity in
estimated inputs changes reported risk rather than treating the fitted model
as exact (Glasserman and Xu, 2014). Here the perturbation set is neither chosen
as a deterministic stress ball nor interpreted as generic model ambiguity. It
is a joint statistical confidence region for the financial primitives, and
the operational question is whether every parameter value it contains lies on
the same side of the spectral boundary.

The spectral map creates an additional difficulty. Simple-eigenvalue
perturbation is locally regular (Kato, 1995), whereas multiple and defective
roots can make spectral functionals nonsmooth or non-Lipschitz (Burke and
Overton, 2001). Econometric inference for directionally differentiable
functionals provides a broader framework for such problems (Fang and Santos,
2019). In the setting studied here, semisimple ties and near-defective matrices
arise inside the estimated impact--risk system and make a pointwise
simple-root interval unreliable as a general regime classifier.

Projection methods translate confidence regions for underlying parameters
into confidence statements for their identified images, including in
partially identified models (Imbens and Manski, 2004; Chernozhukov, Hong and
Tamer, 2007). Their usefulness here depends on the quality of the primitive
region. Weak identification can invalidate a Wald construction before the
spectral functional is evaluated (Dufour, 1997; Andrews and Cheng, 2012), while
dependent-data validity requires control of the long-run covariance (Andrews,
1991; de Jong and Davidson, 2000; Mikusheva, 2007). The analysis consequently
separates primitive identification from propagation through the stability
boundary rather than allowing one problem to conceal the other.

The final link is computational. Interval analysis and verified eigenvalue
bounds can enclose the image of a parameter region, while trust-region
duality supplies sharp bounds for the capacity calculation (Rendl and
Wolkowicz, 1997; Moore, Kearfott and Cloud, 2009; Rump, 2022). These tools make
the direction of numerical error operationally important: incomplete
certification may widen an interval or preserve an unresolved decision, but it
must not manufacture a sign. This separates statistical abstention from a
solver that simply has not finished certifying the relevant range.

Gebbie (2026) gives a complementary hierarchical-causal account in which
higher-level states restrict admissible lower-level laws and their
calendar-time realization. The present analysis begins after one financial
feedback operator has been selected. If the structural assignment or the
event-to-calendar clock is not unique, that ambiguity must first be represented
in the confidence region for the primitive inputs. Likewise, a non-summable
cross-sectional limit of the kind studied by Rodríguez Domínguez (2026c) is
outside the geometric-mixing front end used here; the projection argument can
still be applied once a valid long-memory confidence region is supplied.

\hypertarget{model-inference-capacity}{%
\section*{3. Model, inference, and capacity decisions}\label{model-inference-capacity}}
\addcontentsline{toc}{section}{3. Model, inference, and capacity decisions}

\subsection*{3.1 Boundary and observable primitives}
\addcontentsline{toc}{subsection}{3.1 Boundary and observable primitives}

The stability boundary is inherited from Rodr\'iguez Dom\'inguez (2026b).
The contribution here begins when its inputs are treated as jointly estimated
objects. Write \(\vartheta=(Q,K,c)\), where \(Q\) is post-resilience
conditional risk, \(K\) is temporary cross-impact, and \(c\) is current
effective risk-bearing capacity in compatible units. The feedback operator and
its stability margin are

\[
A(\vartheta)=cQ^{-1}K,
\qquad
m(\vartheta)=\min_{\lambda\in\operatorname{spec}\{A(\vartheta)\}}
\operatorname{Re}\lambda+\frac12 .
\tag{3.1}\label{eq:boundary-margin}
\]

\begin{figure}[t]
\centering
\includegraphics[width=0.94\textwidth]{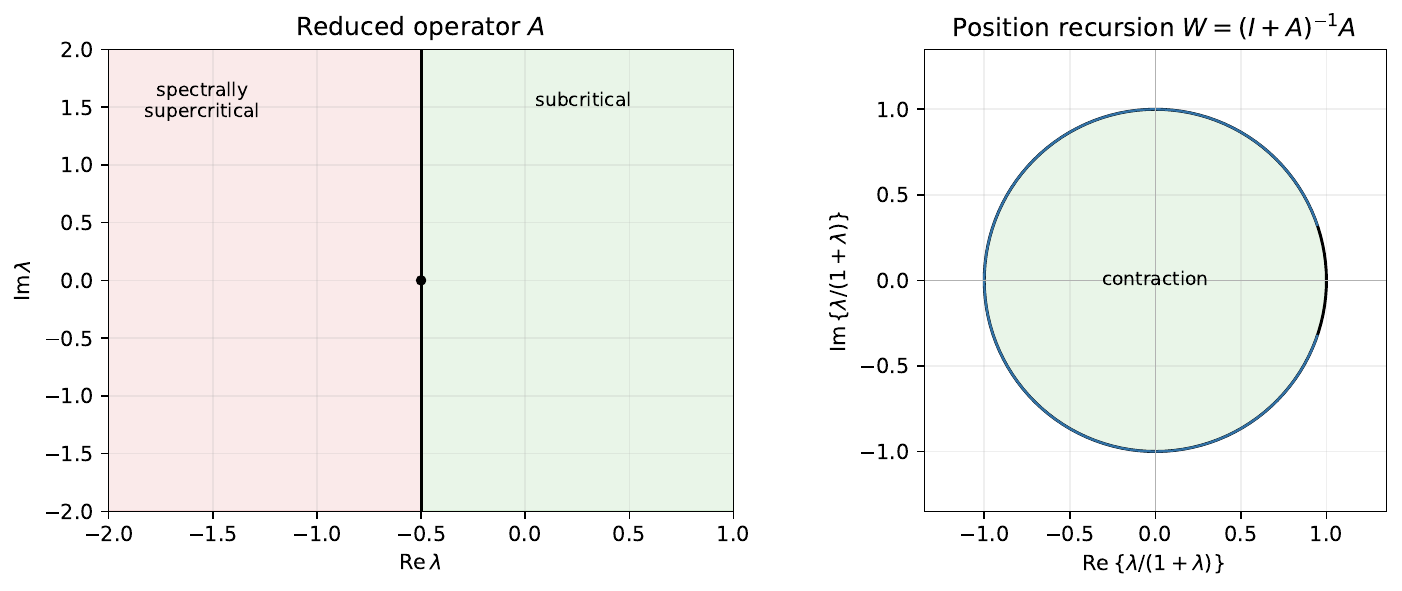}
\caption{The stability boundary under the rational spectral map. The
subcritical half-plane maps to the contraction region of the position
recursion.}
\label{fig:spectral-mapping}
\end{figure}

The regime is subcritical when \(m>0\), on the boundary when \(m=0\), and
spectrally supercritical when \(m<0\). Economic-supercritical language requires
the additional loading and equilibrium conditions in the antecedent model. Its
rational spectral mapping is recorded in Figure \ref{fig:spectral-mapping}; the
derivation is not repeated.

Fix an execution horizon \(h\), a post-resilience horizon \(H>h\), and a
pre-trade state dictionary \(g(Z_t)\). For each block, observe short- and
long-horizon price changes \(R_t^S\) and \(R_t^L\), signed flow \(X_t\), an
excluded flow shifter \(W_t\), the state \(Z_t\), and normalized capacity
\(B_t\). With \(Y_t=R_t^S-R_t^L\), the baseline observation model is

\begin{align}
R_t^L&=M_Qg(Z_t)+U_t,
&E(U_t\mid Z_t)&=0,
&E(U_tU_t'\mid Z_t)&=Q,\nonumber\\
Y_t&=M_Yg(Z_t)+KX_t+\varepsilon_t,
&E(\varepsilon_t\widetilde W_t')&=0,\nonumber\\
X_t&=M_Xg(Z_t)+\Pi W_t+V_t,
&B_t&=c+\eta_t,
&E\eta_t&=0,
\tag{3.2}\label{eq:observation-model}
\end{align}

Collect these observations in
\(O_t=(R_t^S,R_t^L,X_t,W_t,Z_t,B_t)\).
Here \(\widetilde W_t\) is the residual from the linear projection of \(W_t\)
on \(g(Z_t)\). The excluded shifter identifies \(K\) through the usual IV
moment, and its residual covariance with \(X_t\) is assumed uniformly
nonsingular. The horizons and the capacity normalization are part of the
estimand: primitives measured at different horizons or in incompatible units
cannot be inserted into the same operator.

The estimators are deliberately conventional. Conditional risk is the sample
covariance of the long-horizon residuals, cross-impact is linear IV after
residualizing on the declared state, and capacity is the sample mean of
\(B_t\). Their influence functions and the HAC construction are given in
Appendix A. The sufficient conditions A1--A6 impose fixed dimension and
horizons, geometric absolute regularity, adequate uniform moments, the moments
in \eqref{eq:observation-model}, strong IV relevance, interior positive-definite
\(Q\), positive \(c\), and a nonsingular joint long-run covariance estimated by
a pre-specified HAC rule. These are standard strong-identification conditions
(Andrews, 1991; Doukhan, 1994; de Jong and Davidson, 2000), not restrictions on
the distance to the spectral boundary.

Whenever \(\vartheta\) appears in a vector expression below, it denotes the
coordinate vector \((\operatorname{vech}(Q)',\operatorname{vec}(K)',c)'\), of
dimension \(d_\vartheta=p(p+1)/2+p^2+1\). The use of \(\operatorname{vech}\)
for symmetric \(Q\) avoids duplicating its off-diagonal entries; no symmetry is
imposed on \(K\).

\par\medskip\noindent\textbf{Theorem 1 (joint primitive and operator limit).}\par

Under A1--A6, uniformly over the strongly identified class,

\[
\sqrt T\,(\widehat\vartheta-\vartheta)
=T^{-1/2}\sum_{t=1}^T\psi_t+o_P(1)
\ \Rightarrow\ N(0,\Sigma),
\qquad
\widehat\Sigma\overset P\longrightarrow\Sigma .
\tag{3.3}\label{eq:primitive-limit}
\]

All cross-covariances among risk, cross-impact, and capacity are retained in
\(\Sigma\). The operator is differentiable on the parameter interior, with

\[
DA_\vartheta[H_Q,H_K,h_c]
=h_cQ^{-1}K+cQ^{-1}H_K-cQ^{-1}H_QQ^{-1}K,
\tag{3.4}\label{eq:operator-differential}
\]

so \(\widehat A=\widehat c\,\widehat Q^{-1}\widehat K\) also has a joint
root-\(T\) Gaussian limit. Appendix A gives the influence functions, the
vectorized covariance map, and the proof.

This result does not cover weak or drifting instruments, singular conditional
risk, unidentified capacity units, growing dimension, or a data-selected state
dictionary. Those are failures of the primitive experiment. They are distinct
from the spectral nonregularity considered next, which can occur even when
\eqref{eq:primitive-limit} holds uniformly.

\subsection*{3.2 Spectral regularity and collisions}
\addcontentsline{toc}{subsection}{3.2 Spectral regularity and collisions}

Suppose first that a real eigenvalue \(\lambda_0\), or a simple conjugate pair,
uniquely attains the minimum real part in \eqref{eq:boundary-margin} and is
separated from the remaining spectrum. Let \(v_0,w_0\) be corresponding right
and left eigenvectors normalized by \(w_0^*v_0=1\).

\par\medskip\noindent\textbf{Theorem 2 (regular pointwise benchmark).}\par

At such a separated simple root, the margin is differentiable along real
perturbations and

\[
Dm_\vartheta[H_Q,H_K,h_c]
=\operatorname{Re}\!\left\{
w_0^*DA_\vartheta[H_Q,H_K,h_c]v_0
\right\}.
\tag{3.5}\label{eq:regular-derivative}
\]

Consequently, \(\sqrt T\{m(\widehat\vartheta)-m(\vartheta)\}\) is pointwise
Gaussian with the HAC variance of the corresponding scalar influence score.
The usual delta interval is therefore valid at a fixed separated simple root.
Its sensitivity is proportional to the left-right eigenvector condition number
\(\|w_0\|\|v_0\|\), and the approximation is not uniform as the spectral gap
closes or the active mode becomes defective. The proof and the feasible
variance formula are in Appendix A; the perturbation step is standard (Kato,
1995).

The two collision geometries responsible for this loss of uniformity are
different. A semisimple tie preserves the root-\(T\) rate but produces a
nonlinear directional derivative. A defective root can also change the rate
(Kato, 1995; Burke and Overton, 2001; Fang and Santos, 2019).

\par\medskip\noindent\textbf{Proposition 3 (local collision geometry).}\par

At the semisimple boundary \(A_0=-\tfrac12 I_2\),

\[
m'_{A_0}(H)=
\min_{\lambda\in\operatorname{spec}(H)}\operatorname{Re}\lambda .
\tag{3.6}\label{eq:semisimple-derivative}
\]

The map is Hadamard directionally differentiable but not Fr\'echet
differentiable, so a Gaussian operator limit generally becomes a non-Gaussian
minimum-of-roots limit. At the defective boundary

\[
A(a,b)=
\begin{pmatrix}-\tfrac12+a&1\\ b&-\tfrac12+a\end{pmatrix},
\qquad
m(a,b)=a-\sqrt{b_+},
\tag{3.7}\label{eq:jordan-example}
\]

a root-\(T\) perturbation of \(b=0\) induces a \(T^{-1/4}\) margin. Appendix B
proves both collision cases and embeds the perturbations
in the identified IV model rather than adding arbitrary matrix noise.

\begin{figure}[t]
\centering
\includegraphics[width=0.94\textwidth]{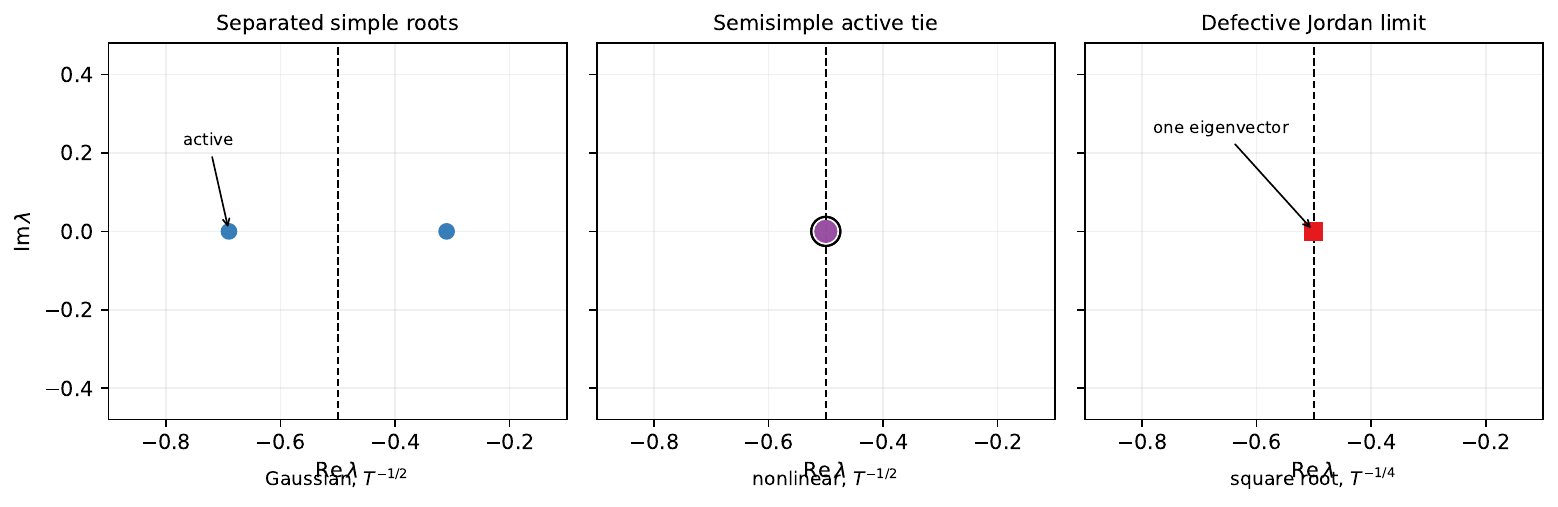}
\caption{Local spectral geometries. Separated simple roots support the
pointwise delta approximation; semisimple ties yield a nonlinear directional
limit; defective roots can change the convergence rate.}
\label{fig:spectral-geometries}
\end{figure}

Figure \ref{fig:spectral-geometries} summarizes the distinction between the
regular, semisimple, and defective local geometries.

The rate change is not merely an exact-boundary pathology. In
\eqref{eq:jordan-example}, take \(b_T=\kappa^2/\sqrt T\). The two true
eigenvalues are distinct at every finite \(T\), but their gap is of the same
order as the square root of the estimation error.

\par\medskip\noindent\textbf{Theorem 3 (nonuniformity of the delta interval).}\par

On any parameter class containing the preceding local sequence, the
simple-root delta interval is not uniformly valid. In the canonical submodel
with \(\sqrt T(\widehat b_T-b_T)\Rightarrow Z\sim N(0,1)\), its limiting
coverage is

\[
P\!\left\{
\left|\kappa-\sqrt{(\kappa^2+Z)_+}\right|
\le \frac{z_{1-\alpha/2}}{2\kappa}
\right\},
\tag{3.8}\label{eq:delta-nonuniformity}
\]

which is not generally \(1-\alpha\). At the conventional 95 per cent level
and \(\kappa=1\), the limit is approximately 84.0 per cent. This counterexample
uses the oracle delta variance, so the failure comes from the smooth spectral
approximation rather than covariance estimation. Appendix B provides the local
sequence and proof. A pretest that selects the smooth formula whenever the
estimated spectrum looks regular does not repair this nonuniformity.

\subsection*{3.3 Projected inference and regime decisions}
\addcontentsline{toc}{subsection}{3.3 Projected inference and regime decisions}

The primitive estimator remains regular across the spectral collisions. The
reported procedure therefore forms a confidence region before applying the
spectral map, following the projection logic used for nonregular and partially
identified parameters (Imbens and Manski, 2004; Chernozhukov, Hong and Tamer,
2007). Let \(\Theta\) impose the declared compact bounds, \(Q\succ0\),
and \(c>0\). With \(q_{d_\vartheta,1-\alpha}\) the relevant chi-square
quantile, define

\[
\mathcal C_T=
\left\{\vartheta\in\Theta:
T(\widehat\vartheta-\vartheta)'
\widehat\Sigma^{-1}(\widehat\vartheta-\vartheta)
\le q_{d_\vartheta,1-\alpha}\right\},
\quad
L_T=\inf_{\vartheta\in\mathcal C_T}m(\vartheta),
\quad
U_T=\sup_{\vartheta\in\mathcal C_T}m(\vartheta).
\tag{3.9}\label{eq:projected-region}
\]

If the primitive covariance check fails or the constrained region is empty,
the procedure reports a primitive-estimation failure; it does not insert an
undeclared ridge or issue a resolved regime.

\par\medskip\noindent\textbf{Theorem 4 (uniform projection and certification).}\par

Under A1--A6,

\[
\liminf_{T\to\infty}\inf_{P\in\mathcal P_T^S}
P\{m(\vartheta(P))\in[L_T,U_T]\}\ge 1-\alpha,
\tag{3.10}\label{eq:uniform-projection}
\]

including at simple roots, semisimple ties, conjugate active pairs, and
defective matrices. If verified computation returns outward bounds
\(\underline L_T\le L_T\) and \(\overline U_T\ge U_T\), the same guarantee
holds for \([\underline L_T,\overline U_T]\). The probability statement follows
directly from coverage of \(\mathcal C_T\); it does not require differentiability
of \(m\). Appendix C gives the uniform Wald argument, the deterministic
projection proof, and the numerical certificate conditions.

The calculation is nonconvex and nonsmooth, so an optimizer's candidate extrema
are insufficient. The reported two-dimensional implementation subdivides the
primitive region and uses verified eigenvalue and tail enclosures. Every
unresolved box can widen the outer interval but cannot create a sign (Moore,
Kearfott and Cloud, 2009; Rump, 2022). The complete algorithm and its stopping
rules are in Appendix C; higher-dimensional
certification would require a different implementation, not a different
statistical theorem.

The regime decision is

\[
\delta_T=
\begin{cases}
\mathrm{SUB},&\underline L_T>0,\\
\mathrm{SUP}_{\mathrm{spec}},&\overline U_T<0,\\
\mathrm{UNR},&\text{otherwise}.
\end{cases}
\tag{3.11}\label{eq:classifier}
\]

\par\medskip\noindent\textbf{Corollary 5 (error and resolution).}\par

The asymptotic probability of a resolved declaration with the wrong sign is at
most \(\alpha\), uniformly over the strongly identified class. If the true
margin remains separated from zero, the primitive region contracts, and the
certificate tolerance vanishes, the correct sign is resolved with probability
tending to one. Along contiguous sequences approaching the boundary, however,
no uniformly valid procedure can be forced to make a binary decision with
vanishing error. Abstention is therefore the necessary counterpart of uniform
control, not an optimization failure. Appendix C supplies the formal statements
and the local impossibility argument, following the logic of Dufour (1997).

\subsection*{3.4 From inference to a capacity action}
\addcontentsline{toc}{subsection}{3.4 From inference to a capacity action}

The same primitive confidence region can restrict future capacity. Current
capacity \(c\) remains part of the estimated state \(\vartheta\); a future
action \(a\ge0\) is chosen after observing the data, and
\(m(Q,K,a)\) denotes \eqref{eq:boundary-margin} with \(c\) replaced by \(a\).
Define the destabilizing intensity

\[
d(Q,K)=\max\!\left\{0,
-\min_{\lambda\in\operatorname{spec}(Q^{-1}K)}
\operatorname{Re}\lambda\right\}.
\tag{3.12}\label{eq:destabilizing-intensity}
\]

For any action \(a\), the future margin is at least
\(\tfrac12-a,d(Q,K)\). Let \(\mathcal C_T^{QK}\) be the projection of
\(\mathcal C_T\) onto \((Q,K)\), and let a verified calculation return

\[
\overline U_{d,T}\ge
\sup_{(Q,K)\in\mathcal C_T^{QK}}d(Q,K).
\tag{3.13}\label{eq:intensity-bound}
\]

Choose an economic buffer \(\eta\in(0,1)\) and a pre-specified bound
\(\rho\ge0\) on proportional implementation overshoot. For finite desired
capacity \(a_{\mathrm{des},T}\), set

\[
a_{\mathrm{plan},T}=
\min\!\left\{a_{\mathrm{des},T},
\frac{1-\eta}{2(1+\rho)\overline U_{d,T}}\right\},
\tag{3.14}\label{eq:safe-capacity}
\]

with the second term interpreted as \(+\infty\) when
\(\overline U_{d,T}=0\). The statistical level \(\alpha\), the economic buffer
\(\eta\), and the execution allowance \(\rho\) solve different problems and
must be reported separately.

\par\medskip\noindent\textbf{Theorem 6 (certified capacity).}\par

Suppose A1--A6 and \eqref{eq:intensity-bound} hold, realized capacity satisfies
\(0\le a_{\mathrm{act},T}\le(1+\rho)a_{\mathrm{plan},T}\), and future
\((Q,K)\) coincide with the estimands covered by \(\mathcal C_T\). Then

\[
\liminf_{T\to\infty}\inf_{P\in\mathcal P_T^S}
P\!\left\{
m(Q(P),K(P),a_{\mathrm{act},T})\ge\frac{\eta}{2}
\right\}
\ge 1-\alpha .
\tag{3.15}\label{eq:capacity-guarantee}
\]

The action in \eqref{eq:safe-capacity} is the largest planned action not
exceeding demand that satisfies this certificate. If the overshoot condition
holds only with probability at least \(1-\beta\), the guarantee becomes
\(1-\alpha-\beta\) without requiring independence. Appendix D gives the proof,
the stochastic-overshoot extension, and the asymmetric loss and drift
interpretations.

This is a conditional capacity statement, not a real-market calibration. It
requires an identified cross-impact moment, compatible capacity units, a
defensible resilience horizon, and stability of \((Q,K)\) over the action
horizon. When any of those empirical conditions fails, the correct output is
no capacity recommendation even if the spectral calculation itself is
available.

\hypertarget{computational-evidence}{%
\section*{4. Computational evidence}\label{computational-evidence}}
\addcontentsline{toc}{section}{4. Computational evidence}

\subsection*{4.1 Design and certification}
\addcontentsline{toc}{subsection}{4.1 Design and certification}

The computational study is designed to separate statistical performance from
numerical certification. It fixes the structural data-generating processes,
identification diagnostics, local spectral geometries, and solver tolerances
before comparing the inferential procedures.

The application must pre-specify:

\begin{enumerate}
\def\labelenumi{\arabic{enumi}.}
\tightlist
\item
  asset universe \(\mathcal U\);
\item
  flow signing and normalization;
\item
  execution horizon \(h\);
\item
  post-resilience horizon \(H\);
\item
  state dictionary \(g(Z_t)\);
\item
  excluded instruments \(W_t\);
\item
  capacity conversion \(\mathcal N\);
\item
  HAC kernel and bandwidth rule;
\item
  the response to a non-positive-definite covariance estimate, including
  whether the entire parameter set is returned;
\item
  primitive bounds, numerical budget and stopping tolerances.
\end{enumerate}

Searching over these choices and reporting the specification with the
most decisive regime classification would invalidate nominal coverage.

At minimum report:

\begin{itemize}
\tightlist
\item
  first-stage singular values of \(\widehat M_{XW}\);
\item
  exclusion-placebo responses before the flow shock;
\item
  the horizon-stability curve \(\widehat K(H)\);
\item
  residual dependence and HAC sensitivity;
\item
  realized HAC bandwidths, minimum covariance eigenvalues and every
  activation of a covariance safeguard;
\item
  covariance eigenvalues of \(\widehat Q\);
\item
  units reconciliation showing \(\widehat c\widehat Q^{-1}\widehat K\)
  is dimensionless;
\item
  sensitivity to information controls and universe matching.
\end{itemize}

For a stratified simulation, Monte Carlo uncertainty must be computed
within design cell and then aggregated using the fixed cell weights.
Cross-cell dispersion is design heterogeneity, not Monte Carlo sampling
error.

Overidentification tests, where available, are diagnostics rather than
proof that the economic exclusion restriction is true.

The baseline can claim estimation of the reflexive boundary only when
all three primitives refer to the same market, universe and horizon. If
signed cross-asset flow, a credible excluded shifter, post-resilience
prices, and capacity data are not jointly available, the appropriate
scope is calibrated or semi-synthetic evidence with the missing data
explicitly.

Existing equity or bond panels that identify only conditional risk and
representation structure are not threshold evidence.

The simulation generates the observable block in \eqref{eq:observation-model}, rather than
adding noise directly to \(Q\) or \(K\). Its stationary recursion is

\[
\begin{aligned}
Z_t&=\phi_ZZ_{t-1}+\nu_t^Z,\\
W_t&=\phi_WW_{t-1}+\nu_t^W,\\
U_t&=\phi_UU_{t-1}+\nu_t^U,
\qquad E[U_tU_t']=Q,\\
V_t&=\phi_VV_{t-1}+\nu_t^V,\\
\eta_t&=\phi_B\eta_{t-1}+\nu_t^B,\\
X_t&=M_Xg(Z_t)+\Pi W_t+V_t,\\
E_t&=L_IV_t+\xi_t,\\
Y_t&=M_Yg(Z_t)+KX_t+E_t,\\
R_t^L&=M_Qg(Z_t)+U_t,\\
B_t&=c+\eta_t.
\end{aligned}
\tag{4.1}
\]

All innovations are Gaussian. The innovations driving \(U_t,V_t\) and
\(\eta_t\) are contemporaneously correlated, while \(W_t\) is
independent of the structural error. Because \(E_t\) depends on \(V_t\),
flow is endogenous in the temporary-price equation, and \(W_t\) shifts
flow through the nonsingular matrix \(\Pi\). Thus OLS is not rescued by
construction, while the IV moment in \eqref{eq:observation-model} remains valid.

This calibration should not be used to claim that nonzero HAC
cross-blocks are empirically necessary. Under the centered Gaussian
design, the independence and parity of the IV score imply zero
population cross-covariances between the \(Q\), \(K\) and \(c\)
influence blocks, even though the underlying innovations are correlated.
The estimator nevertheless retains every sample cross-block, as required
by the general theory. E1--E4 therefore validate the joint
implementation but do not measure the cost of deleting nonzero
population cross-blocks.

Table \ref{tab:design-correspondence} maps the pre-specified experiment to the sufficient class in Section 3.1.

\begin{longtable}[]{@{}
  >{\raggedright\arraybackslash}p{(\columnwidth - 4\tabcolsep) * \real{0.333}}
  >{\raggedright\arraybackslash}p{(\columnwidth - 4\tabcolsep) * \real{0.333}}
  >{\raggedright\arraybackslash}p{(\columnwidth - 4\tabcolsep) * \real{0.333}}@{}}
\caption{Correspondence between assumptions and the pre-specified simulation design.}\label{tab:design-correspondence}\\
\endfirsthead
\toprule\noalign{}
\begin{minipage}[b]{\linewidth}\raggedright
Condition
\end{minipage} & \begin{minipage}[b]{\linewidth}\raggedright
pre-specified design
\end{minipage} & \begin{minipage}[b]{\linewidth}\raggedright
Status and diagnostic
\end{minipage} \\
\midrule\noalign{}
\endhead
\bottomrule\noalign{}
\endlastfoot
A1: dependence & Stable Gaussian Markov recursion with
\(\max|\phi|=0.550\) & Geometrically mixing. Each series starts at zero and
discards 250 observations; the largest marginal covariance discrepancy
is bounded by \(0.550^{500}=1.520\times10^{-130}\). This is covered by the
geometrically coupled initialization extension of A1. \\
A2: moments and design & \(p=2\), \(q=2\), fixed dictionary
\(g(Z)=(1,Z)'\), Gaussian innovations & All required moments exist and
dimensions do not grow. \\
A3: structural moments & \(U_t\) is independent of \(Z_t\); \(W_t\) is
independent of \(E_t\); \(E_t=L_IV_t+\xi_t\) & Conditional-risk and IV
moments hold for the stationary target; endogeneity remains because
\(V_t\) enters both \(X_t\) and \(E_t\). \\
A4: strong IV & \(M_{XW}=\Pi\Omega_W\) & Population minimum singular
value \(0.576\); across the complete design, sample minimum
\(0.303\) and mean \(0.565\). No weak-IV sequence is simulated. \\
A5: interiority & \(\lambda(Q)=(0.795,1.305)\), \(c=1\),
\(\max\|K\|_F=2.189\) & Strictly inside the pre-specified bounds
\(q\in[0.200,3]\), \(c\in[0.400,1.600]\), \(\|K\|_F\le5\). \\
A6: HAC & Bartlett kernel and \(b_T=\lfloor4(T/100)^{2/9}\rfloor\) &
\(b_T=4\) at \(T=250\) and \(6\) at \(T=1000\). All 60,000 HAC matrices
were positive definite and the declared \(10^{-10}\) eigenvalue
safeguard was never activated. \\
\end{longtable}

The stable Gaussian construction and positive idiosyncratic variance in
each score block make the population long-run covariance finite and
nonsingular. The finite-sample diagnostics corroborate, but do not prove
beyond this DGP, the uniform lower bound in A6. The local and Jordan
sequences alter only \(K\); all dependence, moment, identification,
covariance and compactness constants remain common across the triangular
design.

The pre-specified experiment includes six geometries: symmetric
separated, nonnormal diagonalizable, complex active pair, semisimple
tie, near-Jordan transition and exact Jordan. For each
\(T\in\{250,1000\}\), it uses

\[
m_T\in
\left\{
-C,-\frac{0.5}{\sqrt T},0,
\frac{0.5}{\sqrt T},C
\right\},
\qquad C=0.75.
\tag{4.2}
\]

One common \(C=0.75\) was pre-specified for all geometries, signs and
sample sizes; the local and zero sequences were unchanged.

The economic layer fixes \(G(a)=a-a^2/2\), \(a_{\mathrm{des}}=1\),
\(a_{\mathrm{prev}}=0.75\), \(\eta=0.10\) and \(\rho=0.05\). It compares
low, baseline and high convention-loss calibrations without claiming
regret dominance.

The numerical stopping rule separates three logically different
outcomes. First, an exterior interval wholly above or below zero
certifies the regime. Second, a nonpositive feasible margin and a
nonnegative feasible margin certify that the projected identified set
straddles zero; this is statistical abstention, not numerical failure.
Third, the remaining numerical uncertainty in the capacity decision is
measured directly as the gap between the safe capacity based on the
exterior intensity bound and the capacity implied by the best certified
feasible intensity. The outer interval and capacity guarantee remain
valid under every stopping outcome.

The implementation uses a ball-aware certificate based on the exact
two-dimensional stability criterion. For
\(D\ge0\), let \(C_D=Q^{-1}K+DI_2\). The real parts of both eigenvalues
of \(C_D\) are nonnegative if and only if

\[
\operatorname{tr}(C_D)\ge0,
\qquad
\det(C_D)\ge0.
\tag{4.3}
\]

Because \(Q\succ0\), these conditions are equivalent to

\[
N_{\mathrm{tr}}(Q,K)+2D\det(Q)\ge0,
\qquad
\det(K+DQ)\ge0,
\tag{4.4}
\]

where

\[
N_{\mathrm{tr}}(Q,K)
=q_{22}k_{11}+q_{11}k_{22}-q_{12}(k_{21}+k_{12}).
\tag{4.5}
\]

Under the affine Wald representation \(\vartheta(z)=\widehat\vartheta+Lz\),
both expressions in (4.4) are quadratic in \(z\). For a quadratic
\(q(z)=q_0+\ell'z+z'Hz\) and any \(\lambda\ge0\) such that
\(H+\lambda I\succ0\),

\[
\inf_{\|z\|\le r}q(z)
\ge
q_0-\lambda r^2-\frac14\ell'(H+\lambda I)^{-1}\ell.
\tag{4.6}
\]

Every use of (4.6) is evaluated with outward interval arithmetic;
interval Cholesky certifies \(H+\lambda I\succ0\) and encloses the
inverse quadratic form. Bisection returns a verified \(D_U\) satisfying
\(d(\vartheta)\le D_U\) on the complete Wald ball. The same trace
polynomial supplies a uniform \(D_L\le d(\vartheta)\) when available.
Therefore

\[
\underline m
\ge \frac12-c_U D_U,
\qquad
\overline m
\le \frac12-c_L D_L,
\tag{4.7}
\]

The upper-margin inequality in (4.7) is used only when the trace
certificate also establishes

\[
\min\operatorname{Re}\operatorname{spec}(Q^{-1}K)
\le-D_L\le0
\]

uniformly on the Wald ball. On that certified set,
\(d=-\min\operatorname{Re}\operatorname{spec}(Q^{-1}K)\ge D_L\), and
hence \(m=1/2-cd\le1/2-c_LD_L\). Without this sign condition, \(D_L\) is
not used to upper-bound the margin.

These certified inequalities sharpen classification and capacity without
changing the confidence region. A symmetric-part robust PSD certificate
is retained only as a valid fallback.

\subsection*{4.2 Results, interpretation, and limits}
\addcontentsline{toc}{subsection}{4.2 Results, interpretation, and limits}

The results are reported first for coverage and regime resolution, then for
capacity decisions and regret. The discussion that follows states the scope
of those results and the identification and normalization limitations that
remain for an empirical application.

The reported design fixes 1,000 replications in each of the 60 cells, so
the worst-case cell-level Monte Carlo standard error for a binary statistic is
at most \(0.5/\sqrt{1000}=0.016\). Replications are indexed uniquely by sample
size, spectral geometry, boundary distance, and replication number. Both the
primitive-region and spectral certificates are checked before a replication
enters the aggregates.

Because the grid is fixed and equally weighted, the global estimand is
the average of the 60 cell means. For any replication statistic \(H\),
its Monte Carlo standard error is therefore computed as

\[
\widehat{\operatorname{se}}_{\mathrm{MC}}(\bar H)
=
\left\{
\frac1{J^2}\sum_{j=1}^J\frac{s_j^2}{n_j}
\right\}^{1/2},
\qquad J=60,
\quad n_j=1000,
\tag{4.8}
\]

where \(s_j^2\) is the within-cell sample variance. A standard deviation
computed after pooling all raw rows would mix cross-cell design
dispersion with simulation noise and is not used below.

Equation (4.8) treats distinct design keys as independent. The reported
run contains two disclosed collisions in its 32-bit seed map, both
across rather than within design cells. For a statistic bounded in
\([0,1]\), the largest possible additional standard-error component from
two perfectly correlated pairs is \(1/60000=0.002\) percentage points.
The collisions are therefore retained and disclosed rather than removed
by selective rerunning; future experiments require a collision-free seed
map.

The full Monte Carlo is complete: 60,000 of 60,000 inference
replications succeed, with no failed primitive estimates and no rejected
spectral certificates. The associated financial-policy comparison contains
1,260,000 evaluations across the declared loss and decision rules.

The completed experiment records coverage of the projected spectral margin,
not membership of the true full primitive vector in the affine Wald region.
A primitive-region coverage percentage is therefore not reported. This
measurement limitation does not change the analytical coverage result, but it
narrows what the simulation directly verifies.

\begin{figure}[t]
\centering
\includegraphics[width=0.94\textwidth]{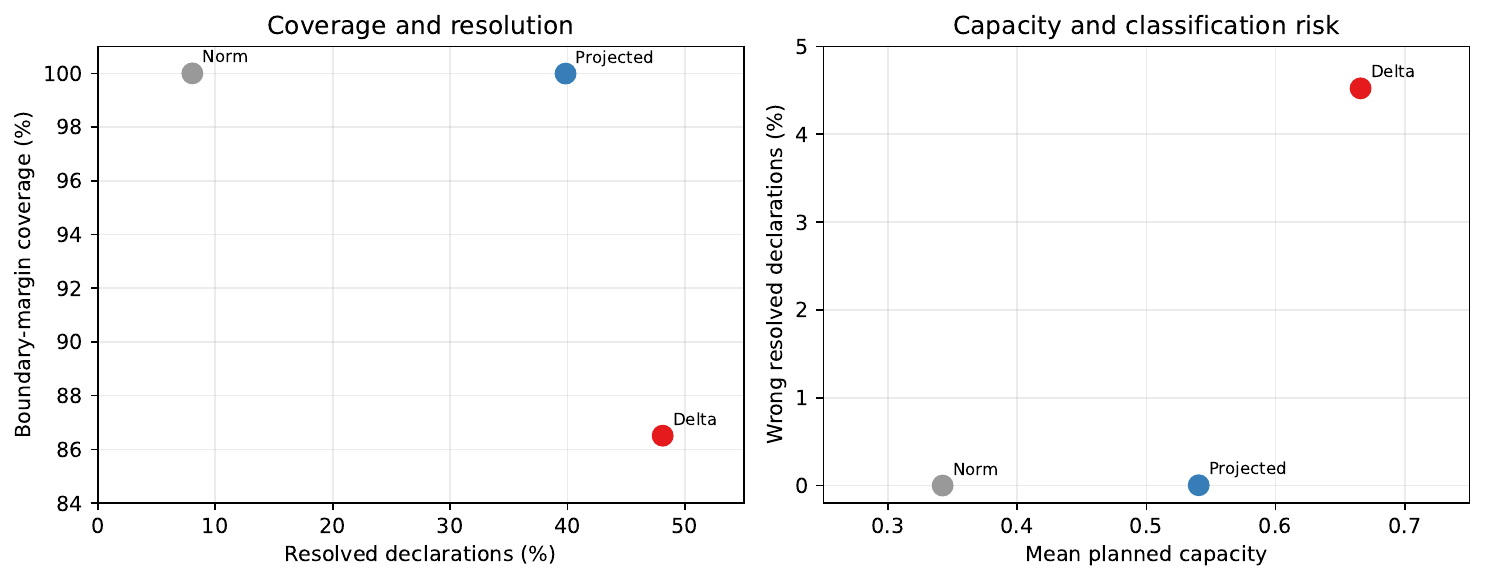}
\caption{Finite-design tradeoffs. The global averages summarize the
fixed equal weighting of the 60 design cells and are not population
frequencies. Cell-level results by boundary distance are reported below.}
\label{fig:tradeoff}
\end{figure}

Figure \ref{fig:tradeoff} summarizes the coverage--resolution--capacity tradeoff.

Table \ref{tab:global-mc} reports the global Monte Carlo comparison.

\begin{longtable}[]{@{}
  >{\raggedright\arraybackslash}p{(\columnwidth - 12\tabcolsep) * \real{0.111}}
  >{\raggedleft\arraybackslash}p{(\columnwidth - 12\tabcolsep) * \real{0.148}}
  >{\raggedleft\arraybackslash}p{(\columnwidth - 12\tabcolsep) * \real{0.148}}
  >{\raggedleft\arraybackslash}p{(\columnwidth - 12\tabcolsep) * \real{0.148}}
  >{\raggedleft\arraybackslash}p{(\columnwidth - 12\tabcolsep) * \real{0.148}}
  >{\raggedleft\arraybackslash}p{(\columnwidth - 12\tabcolsep) * \real{0.148}}
  >{\raggedleft\arraybackslash}p{(\columnwidth - 12\tabcolsep) * \real{0.148}}@{}}
\caption{Global Monte Carlo comparison of delta and projected inference.}\label{tab:global-mc}\\
\endfirsthead
\toprule\noalign{}
\begin{minipage}[b]{\linewidth}\raggedright
Method
\end{minipage} & \begin{minipage}[b]{\linewidth}\raggedleft
Coverage
\end{minipage} & \begin{minipage}[b]{\linewidth}\raggedleft
MC SE
\end{minipage} & \begin{minipage}[b]{\linewidth}\raggedleft
Wrong resolved
\end{minipage} & \begin{minipage}[b]{\linewidth}\raggedleft
MC SE
\end{minipage} & \begin{minipage}[b]{\linewidth}\raggedleft
Unresolved
\end{minipage} & \begin{minipage}[b]{\linewidth}\raggedleft
MC SE
\end{minipage} \\
\midrule\noalign{}
\endhead
\bottomrule\noalign{}
\endlastfoot
Pointwise delta & 86.510\% & 0.133 pp & 4.523\% & 0.077 pp &
51.900\% & 0.111 pp \\
Deterministic norm & 100.000\% & 0.000 pp & 0.000\% & 0.000 pp &
91.948\% & 0.040 pp \\
Projected certified envelope & 99.992\% & 0.004 pp & 0.003\% & 0.002
pp & 60.170\% & 0.018 pp \\
\end{longtable}

The plug-in classifier is wrong in 34.863\% of the full design. The
projected procedure therefore occupies the intended middle ground.
Relative to the deterministic norm bound, it resolves far more cells
while retaining near-complete empirical coverage. Relative to pointwise
delta inference, it materially increases coverage and reduces wrong
resolved declarations from 4.523\% to 0.003\%.

Table \ref{tab:distance-results} reports the results by distance from the boundary. The two projected wrong resolved declarations both occur at the exact
boundary and coincide with projected-envelope coverage misses. There are
five projected coverage misses in total. Counts are reported because
rounding the wrong resolved rate to zero would conceal these cases.

\begin{longtable}[]{@{}
  >{\raggedright\arraybackslash}p{(\columnwidth - 10\tabcolsep) * \real{0.130}}
  >{\raggedleft\arraybackslash}p{(\columnwidth - 10\tabcolsep) * \real{0.174}}
  >{\raggedleft\arraybackslash}p{(\columnwidth - 10\tabcolsep) * \real{0.174}}
  >{\raggedleft\arraybackslash}p{(\columnwidth - 10\tabcolsep) * \real{0.174}}
  >{\raggedleft\arraybackslash}p{(\columnwidth - 10\tabcolsep) * \real{0.174}}
  >{\raggedleft\arraybackslash}p{(\columnwidth - 10\tabcolsep) * \real{0.174}}@{}}
\caption{Projected-inference results by distance from the stability boundary.}\label{tab:distance-results}\\
\endfirsthead
\toprule\noalign{}
\begin{minipage}[b]{\linewidth}\raggedright
Distance
\end{minipage} & \begin{minipage}[b]{\linewidth}\raggedleft
Projected coverage
\end{minipage} & \begin{minipage}[b]{\linewidth}\raggedleft
Wrong resolved
\end{minipage} & \begin{minipage}[b]{\linewidth}\raggedleft
Unresolved
\end{minipage} & \begin{minipage}[b]{\linewidth}\raggedleft
Classification complete
\end{minipage} & \begin{minipage}[b]{\linewidth}\raggedleft
Capacity precise
\end{minipage} \\
\midrule\noalign{}
\endhead
\bottomrule\noalign{}
\endlastfoot
Fixed supercritical & 99.992\% & 0.000\% & 0.000\% & 100.000\% &
100.000\% \\
Local supercritical & 99.992\% & 0.000\% & 99.933\% & 99.942\% &
99.900\% \\
Boundary & 99.983\% & 0.017\% & 99.983\% & 100.000\% & 99.767\% \\
Local subcritical & 99.992\% & 0.000\% & 99.975\% & 100.000\% &
99.733\% \\
Fixed subcritical & 100.000\% & 0.000\% & 0.958\% & 99.817\% &
99.933\% \\
\end{longtable}

The local and exact-boundary nonresolution rates are statistical
abstention, not failed forced classification. Fixed-distance cells
resolve almost completely. This ordering is consistent with the uniform
boundary theory; the simulation illustrates rather than proves the
asymptotic resolution statements.

Table \ref{tab:geometry-results} reorganizes the same reported output by spectral geometry and sample size. Each main percentage averages equally over the five
distance cells, with 1,000 replications per cell. The last column
isolates the two fixed-distance cells so that difficulty caused by
geometry is not hidden by the three local or boundary cells, which are
designed to abstain.

\begin{longtable}[]{@{}
  >{\raggedleft\arraybackslash}p{(\columnwidth - 10\tabcolsep) * \real{0.174}}
  >{\raggedright\arraybackslash}p{(\columnwidth - 10\tabcolsep) * \real{0.130}}
  >{\raggedleft\arraybackslash}p{(\columnwidth - 10\tabcolsep) * \real{0.174}}
  >{\raggedleft\arraybackslash}p{(\columnwidth - 10\tabcolsep) * \real{0.174}}
  >{\raggedleft\arraybackslash}p{(\columnwidth - 10\tabcolsep) * \real{0.174}}
  >{\raggedleft\arraybackslash}p{(\columnwidth - 10\tabcolsep) * \real{0.174}}@{}}
\caption{Projected-inference results by spectral geometry and sample size.}\label{tab:geometry-results}\\
\endfirsthead
\toprule\noalign{}
\begin{minipage}[b]{\linewidth}\raggedleft
\(T\)
\end{minipage} & \begin{minipage}[b]{\linewidth}\raggedright
Spectral geometry
\end{minipage} & \begin{minipage}[b]{\linewidth}\raggedleft
Coverage
\end{minipage} & \begin{minipage}[b]{\linewidth}\raggedleft
Wrong resolved
\end{minipage} & \begin{minipage}[b]{\linewidth}\raggedleft
Unresolved, all distances
\end{minipage} & \begin{minipage}[b]{\linewidth}\raggedleft
Unresolved, fixed distances
\end{minipage} \\
\midrule\noalign{}
\endhead
\bottomrule\noalign{}
\endlastfoot
250 & Symmetric separated & 100.000\% & 0.000\% & 60.000\% &
0.000\% \\
250 & Nonnormal diagonalizable & 100.000\% & 0.000\% & 60.240\% &
0.600\% \\
250 & Complex active pair & 99.980\% & 0.000\% & 59.980\% &
0.000\% \\
250 & Semisimple tie & 100.000\% & 0.000\% & 59.940\% & 0.000\% \\
250 & Near-Jordan transition & 99.980\% & 0.000\% & 60.320\% &
0.850\% \\
250 & Exact Jordan & 100.000\% & 0.000\% & 61.700\% & 4.300\% \\
1000 & Symmetric separated & 99.980\% & 0.020\% & 59.940\% &
0.000\% \\
1000 & Nonnormal diagonalizable & 100.000\% & 0.000\% & 60.000\% &
0.000\% \\
1000 & Complex active pair & 99.980\% & 0.000\% & 59.980\% &
0.000\% \\
1000 & Semisimple tie & 100.000\% & 0.000\% & 59.980\% & 0.000\% \\
1000 & Near-Jordan transition & 100.000\% & 0.000\% & 60.000\% &
0.000\% \\
1000 & Exact Jordan & 99.980\% & 0.020\% & 59.960\% & 0.000\% \\
\end{longtable}

At \(T=250\), the fixed-distance unresolved rate rises from zero in the
symmetric, complex and semisimple designs to 0.60\% under nonnormality,
0.85\% near the Jordan transition and 4.30\% at the exact Jordan block.
At \(T=1000\), every fixed-distance geometry resolves. The two wrong
resolved declarations are the already disclosed exact-boundary misses in
the symmetric and exact-Jordan rows; they are not fixed-distance
failures.

The solver terminates adaptively in 59,892 runs and exhausts the
pre-specified 400-box budget in 108 runs. Classification completes in
99.952\% of replications and the capacity-precision target is met in
99.867\%. Twenty-nine runs do not complete classification, 80 do not
attain the capacity-action tolerance, and one misses both components.
The mean capacity-action gap is 0.003, while the maximum is
0.14. Budget-exhausted outputs remain outward-certified but are
not described as having met the adaptive precision criterion.

The exact two-dimensional Routh--Hurwitz ball certificate is used in
58,895 replications; the conservative symmetric-part fallback is used in
1,105. There are no primitive-estimation failures and no retained
fallback-active boxes. The minimum certified guaranteed margin is
0.05, equal to the target 0.05 up to outward numerical
rounding. Two 32-bit seed collisions occur among the 60,000 design keys.
They are reported in Appendix F; the affected outcomes are retained rather
than selectively rerun.

Table \ref{tab:capacity-results} gives the financial comparison under the baseline loss convention.

\begin{longtable}[]{@{}
  >{\raggedright\arraybackslash}p{(\columnwidth - 8\tabcolsep) * \real{0.158}}
  >{\raggedleft\arraybackslash}p{(\columnwidth - 8\tabcolsep) * \real{0.2105}}
  >{\raggedleft\arraybackslash}p{(\columnwidth - 8\tabcolsep) * \real{0.2105}}
  >{\raggedleft\arraybackslash}p{(\columnwidth - 8\tabcolsep) * \real{0.2105}}
  >{\raggedleft\arraybackslash}p{(\columnwidth - 8\tabcolsep) * \real{0.2105}}@{}}
\caption{Capacity and regret under the reported decision rules.}\label{tab:capacity-results}\\
\endfirsthead
\toprule\noalign{}
\begin{minipage}[b]{\linewidth}\raggedright
Policy
\end{minipage} & \begin{minipage}[b]{\linewidth}\raggedleft
Planned capacity
\end{minipage} & \begin{minipage}[b]{\linewidth}\raggedleft
Violation rate
\end{minipage} & \begin{minipage}[b]{\linewidth}\raggedleft
Mean regret
\end{minipage} & \begin{minipage}[b]{\linewidth}\raggedleft
Regret MC SE
\end{minipage} \\
\midrule\noalign{}
\endhead
\bottomrule\noalign{}
\endlastfoot
Oracle & 0.857 & 0.000\% & 0.000 & 0.000 \\
Projected safe & 0.541 & 0.000\% & 0.104 & 0.000 \\
Deterministic norm & 0.342 & 0.000\% & 0.196 & 0.000 \\
Full withdrawal & 0.000 & 0.000\% & 0.489 & 0.000 \\
Pointwise delta & 0.666 & 1.758\% & 0.066 & 0.000 \\
Plug-in & 0.758 & 6.233\% & 0.083 & 0.001 \\
Desired unprotected & 1.000 & 58.333\% & 1.673 & 0.000 \\
\end{longtable}

The projected-safe policy records zero boundary violations in every
design cell and improves capacity action and regret relative to the
other non-oracle zero-violation policies. It does not dominate policies
that accept positive violation risk under the baseline or low
convention-loss calibration. Under the high convention-loss calibration,
projected-safe mean regret is 0.107, below pointwise delta at
0.120 and plug-in at 0.271. The evidence is therefore a
safety-performance tradeoff, not unconditional regret dominance.

The baseline experiment holds the primitive information design fixed. Two
operational variants ask how the procedure behaves when information quality
or computational effort changes. The first reconstructs
six representative rolling conditional-risk matrices from the observed equity
and bond panels, combines each with symmetric, nonnormal, and exact-Jordan
cross-impact geometries, and varies only the uncertainty carried by the risk
estimate. Cross-impact and capacity remain imposed, so this is an
observed-risk numerical stress rather than structural market validation.
Across 522 certified runs there are no rejected numerical certificates and no
wrong fixed-distance classifications.

\begin{figure}[t]
\centering
\includegraphics[width=0.94\textwidth]{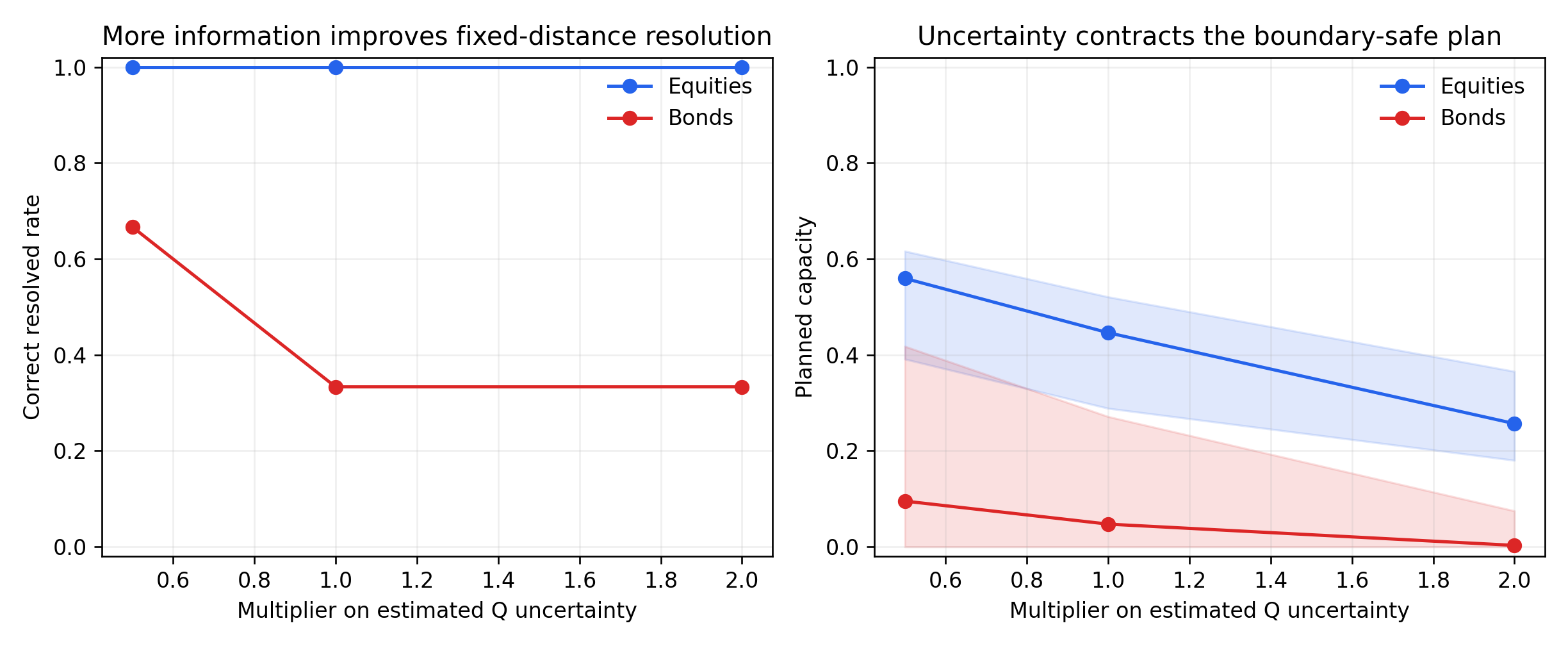}
\caption{Observed-risk information stress. Left: correct resolution of fixed
regimes as uncertainty in the conditional-risk estimate changes. Right:
planned capacity at the exact boundary under the same information changes.
Cross-impact and capacity are imposed, so the figure evaluates the decision
procedure rather than validating a market boundary.}
\label{fig:information-capacity}
\end{figure}

Figure \ref{fig:information-capacity} shows how information quality affects
resolution and action. All fixed-distance equity cases resolve at each tested
uncertainty multiplier. The corresponding bond resolution rate falls from
0.667 at half the baseline risk uncertainty to 0.333 at the baseline and
double-uncertainty settings. At the exact boundary, median planned capacity
falls from 0.559 to 0.257 for equities and from 0.095 to 0.003 for bonds as the
uncertainty multiplier rises from 0.500 to 2.000. The appropriate operational
response to weak information is therefore less capacity and more abstention,
not a more decisive plug-in label.

\begin{figure}[t]
\centering
\includegraphics[width=0.94\textwidth]{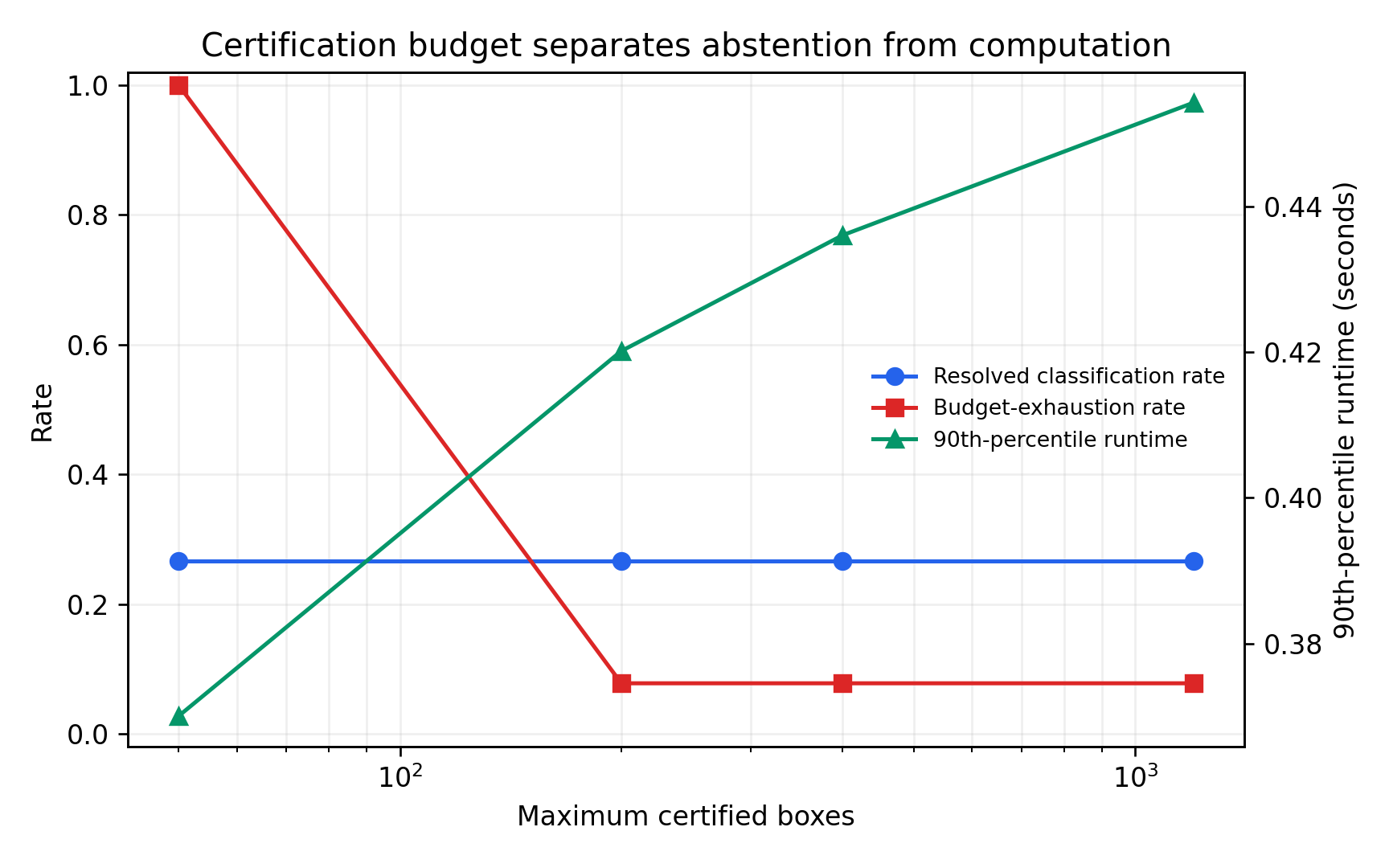}
\caption{Certification-budget stress. The budget-exhaustion rate drops after
the smallest budget, while regime resolution does not change. This separates
computational incompleteness from statistical abstention.}
\label{fig:certification-budget}
\end{figure}

The second variant changes the branch-and-bound budget while leaving the
confidence region unchanged. Figure \ref{fig:certification-budget} shows that
raising the budget from 50 to 200 boxes reduces budget exhaustion from 1.000
to 0.078, but fixed-regime resolution remains 0.267 and does not improve at
budgets of 400 or 1,200. Once the modest computational threshold is reached,
the remaining nonresolution is statistical. Spending more computation cannot
replace information, and an unresolved label should not be overridden by a
larger search budget.

The completed experiment supports finite-sample claims only for the
pre-specified simulated design. It does not establish universal
coverage, zero wrong-classification probability, universal adaptive
completion, real-market capacity validity, high-dimensional
certification, unconditional regret dominance or causal identification
from observed market data.

The interpretation of these results is limited by the maintained observation, identification, and normalization conditions.

If information is incorporated gradually beyond \(H\), then
\(R_t^S-R_t^L\) can mix transient impact with delayed information. The
response is not to assume a longer \(H\) mechanically. The paper must
show horizon stability and use an excluded source of flow variation
whose effect on fundamental news is defensible.

Theorems 1--2 assume \(\sigma_{\min}(M_{XW})\) is bounded away from
zero. If the first stage is weak, \(\widehat K\) is not regular and the
primitive Wald region is invalid.

The definition of effective capacity in Section 3.1 forces the application to
disclose the transformation from balance sheet and holdings into the
equilibrium multiplier. Without
a credible \(\mathcal N\), the paper can estimate \(A\) only up to scale
or can study \(cK\) directly, but cannot issue a capacity
recommendation.

Theorem 2 is classical eigenvalue perturbation composed with the joint
primitive estimator. Its role is to provide a transparent pointwise
benchmark for the nonregular analysis.

The assumptions are observable in the sense that each maps to a timing
choice, diagnostic, instrument claim, normalization, or spectral
statistic. Hiding them would make the paper shorter but weaker.

\subsection*{4.3 Operational capacity decisions}
\addcontentsline{toc}{subsection}{4.3 Operational capacity decisions}

The protocol converts the joint estimate of \((Q,K,c)\), its long-run
covariance, instrument specification, capacity normalization, and action
horizon into a certified margin, classifier, intensity bound, and capacity
plan. Table \ref{tab:operational-protocol} records the permitted actions.

\begin{longtable}[]{@{}p{0.18\columnwidth}p{0.31\columnwidth}p{0.43\columnwidth}@{}}
\caption{Operational interpretation of the certified output.}\label{tab:operational-protocol}\\
\toprule
Output & Statistical meaning & Permitted operational use \\
\midrule
\endfirsthead
\toprule
Output & Statistical meaning & Permitted operational use \\
\midrule
\endhead
\bottomrule
\endlastfoot
\(\mathrm{SUB}\) & The projected margin is strictly positive. & Implement no more than the smaller of desired and certified safe capacity. The label does not authorize deployment above the one-sided limit. \\
\(\mathrm{SUP}_{\mathrm{spec}}\) & The projected margin is strictly negative. & Cap future deployment and initiate a separate reduction analysis. The classifier does not solve an optimal unwind. \\
\(\mathrm{UNR}\) & The confidence set crosses the boundary or certified computation cannot sharpen it. & Abstain from the regime claim and use only the one-sided conditional capacity restriction. \\
Failed empirical condition & Identification, normalization, resilience horizon, or parameter stability is not defended. & Issue no capacity recommendation, even if a spectral calculation can be produced. \\
\end{longtable}

Application of the protocol requires explicit declarations on identification,
normalization, resilience, and the stability of \(Q,K\) over the action
horizon. Desired, certified, planned, and realized capacity must be recorded
separately, together with any abstention or refusal. In the worked example the
margin interval crosses zero, so the decision is \(\mathrm{UNR}\) and desired
capacity falls from \(1.000\) to the conditional plan \(0.441\), retaining
margin \(0.050\). These values demonstrate the decision map, not a limit for a
real portfolio.

\hypertarget{conclusion}{%
\section*{5. Conclusion}\label{conclusion}}
\addcontentsline{toc}{section}{5. Conclusion}

This paper joins estimation of conditional risk, cross-impact, and capacity
to a regime decision at a spectral stability boundary. The joint observation
model retains dependence and cross-covariances that would be lost if the three
inputs were estimated and assessed separately. When the spectrum is regular,
the familiar pointwise approximation remains an informative benchmark. Near
semisimple or defective collisions, projection of the primitive confidence
region provides a regime interval that can abstain rather than convert local
instability into a false declaration. The same confidence region then limits
capacity, so the statistical evidence used for classification is consistent
with the evidence used for action.

The structural experiment supports the predicted coverage--resolution
tradeoff. Across 60,000 replications, the projected-safe rule records no
observed boundary violations within the stated design and preserves more
capacity, with lower regret, than the other non-oracle policies that also
record no violations. These comparisons are properties of the experimental
design, not estimates of real-market performance.

The observed-risk variants add the practical distinction the baseline design
cannot show: poor information lowers both resolution and allowable capacity,
whereas additional computation stops helping once the certification budget is
large enough. This distinction makes abstention interpretable as either a
computational warning or a statistical decision, rather than a generic solver
failure.

These conclusions do not constitute a real-market capacity claim. An empirical application must still defend the normalization
\(\mathcal N\), the exclusion restriction for \(W_t\), the resilience
horizon and the stability of \(Q,K\) over the capacity action horizon.
Weak primitive identification would require a different primitive
confidence set before projection. Higher-dimensional sharp certification
and robustness to failed identification are separate extensions, not
missing components of the present result.

E1--E4 is pre-specified without tuning after observing the
complete-design results. No selective rerun, solver-budget enlargement
or replacement of the five coverage misses is used in the reported
evidence.

\begin{center}\rule{0.5\linewidth}{0.5pt}\end{center}

\clearpage
\hypertarget{app-appendix-a.-primitive-regularity-and-asymptotic-linearity}{%
\section*{Appendix A. Primitive regularity and asymptotic
linearity}\label{app:appendix-a.-primitive-regularity-and-asymptotic-linearity}}
\addcontentsline{toc}{section}{Appendix A. Primitive regularity and
asymptotic linearity}

\hypertarget{app-baseline-parameter-class}{%
\subsection*{Baseline parameter
class}\label{app:baseline-parameter-class}}
\addcontentsline{toc}{subsection}{Baseline parameter class}

Let \(\mathcal P^S\) be the strongly identified class used by the
reported Wald procedure. The dimensions \(p,q\), the horizons \(h,H\),
and the dictionary \(g\) are fixed. Every \(P\in\mathcal P^S\) satisfies
the following conditions with the same finite constants.

\textbf{A1. Uniform dependence.} \(\{O_t\}\) is strictly stationary and
absolutely regular. For constants \(C_\beta<\infty\) and
\(0<\rho_\beta<1\),

\[
\sup_{P\in\mathcal P^S}\beta_P(j)
\le C_\beta\rho_\beta^j,
\qquad j\ge1.
\]

Non-overlapping blocks are not required, but the overlap generated by
the fixed horizon \(H\) is part of this dependence sequence. The
assumption is stronger than necessary, but places the score array inside
a standard mixing-CLT framework (Doukhan, 1994).

The same conclusions allow an initialized process \(O_{t,T}\) coupled to
a stationary version \(O_t^*\) when, for a common \(C_0<\infty\) and
\(0<\rho_0<1\),

\[
\sup_{P\in\mathcal P^S}
E_P\|O_{t,T}-O_t^*\|
\le C_0\rho_0^{r_T+t},
\qquad
T^{-1/2}\sum_{t=1}^T\rho_0^{r_T+t}\to0,
\]

where \(r_T\) is the discarded burn-in. This permits a documented
geometrically vanishing initialization transient but not an unrestricted
structural break.

\textbf{A2. Uniform moments and state design.} For some \(\delta>0\),
with

\[
V_t^*=(R_t^L,Y_t,X_t,W_t,g_t,B_t),
\]

\[
\sup_{P\in\mathcal P^S}
E_P\|V_t^*\|^{8+\delta}<\infty,
\qquad
\inf_{P\in\mathcal P^S}
\lambda_{\min}\{E_P[g_tg_t']\}>0,
\]

and the corresponding population projection coefficients are uniformly
bounded. These conditions give uniformly bounded \(4+\delta/2\) moments
for the quadratic risk score and the IV product score, which are
sufficient for the score CLT and the retained-lag HAC stochastic bound
used below.

\textbf{A3. Structural moments.} The conditional mean, conditional
covariance, IV, and capacity restrictions in
\eqref{eq:observation-model} hold.

\textbf{A4. Strong IV relevance.} With
\(M_{XW}=E(\widetilde X_t\widetilde W_t')\),
\(\sigma_{\min}(M_{XW})\ge\underline\kappa_W>0\) for every
\(P\in\mathcal P^S\).

\textbf{A5. Primitive compactness and interiority.}

\[
0<\underline q
\le\lambda_{\min}(Q)
\le\lambda_{\max}(Q)
\le\bar q,
\quad
\|K\|\le\bar k,
\quad
0<\underline c\le c\le\bar c.
\tag{A.1}
\]

The true primitive lies in a fixed interior subset of these bounds. The
restrictions therefore stabilize inversion and optimization; they are
not locally binding devices used to manufacture coverage.

\textbf{A6. Score covariance and fixed HAC rule.} The joint long-run
covariance \(\Sigma(P)\) in (A.6) obeys

\[
0<\underline\sigma
\le\lambda_{\min}\{\Sigma(P)\}
\le\lambda_{\max}\{\Sigma(P)\}
\le\bar\sigma<\infty
\]

uniformly over \(\mathcal P^S\). Let \(k\) be a fixed bounded,
symmetric, Lipschitz kernel supported on \([-1,1]\), with \(k(0)=1\),
and let the pre-specified bandwidth satisfy

\[
b_T\to\infty,
\qquad
b_T=o(T^{1/2}).
\]

Using the plug-in influence scores and
\(\bar{\widehat\psi}=T^{-1}\sum_t\widehat\psi_t\), define

\[
\widehat\Gamma_\ell
=
T^{-1}\sum_{t=\ell+1}^T
(\widehat\psi_t-\bar{\widehat\psi})
(\widehat\psi_{t-\ell}-\bar{\widehat\psi})',
\qquad
\widehat\Sigma
=
\widehat\Gamma_0
+\sum_{\ell=1}^{b_T}
k\{\ell/(b_T+1)\}
(\widehat\Gamma_\ell+\widehat\Gamma_\ell').
\]

The kernel and bandwidth sequence are pre-specified before examining the
regime results. A numerical ridge may be used only as a declared
finite-sample failure safeguard; it is not part of the asymptotic
covariance theorem.

These are strong sufficient conditions for the
fixed-dimensional regular primitive experiment. They make the uniformity
claim auditable. Slower mixing, growing horizons, data-selected
dictionaries, high dimension and weak-IV sequences require separate
arguments and are not included in \(\mathcal P^S\).

The estimators used in the paper are the sample analogues of these moments.
Writing \(g_t=g(Z_t)\), let

\[
\widehat M_Q=
\left(T^{-1}\sum_tR_t^Lg_t'\right)
\left(T^{-1}\sum_tg_tg_t'\right)^{-1},
\qquad
\widehat Q=T^{-1}\sum_t\widehat U_t\widehat U_t',
\quad
\widehat U_t=R_t^L-\widehat M_Qg_t.
\]

After linearly residualizing \(Y_t,X_t,W_t\) on \(g_t\), define

\[
\widehat K=\widehat M_{YW}\widehat M_{XW}^{-1},
\qquad
\widehat M_{YW}=T^{-1}\sum_t
\widehat{\widetilde Y}_t\widehat{\widetilde W}_t',
\qquad
\widehat M_{XW}=T^{-1}\sum_t
\widehat{\widetilde X}_t\widehat{\widetilde W}_t',
\]

and \(\widehat c=T^{-1}\sum_tB_t\). With additional instruments,
\(\widehat K\) is replaced by the corresponding GMM estimator and sandwich
influence function. The displayed just-identified form keeps the financial
moment explicit.

\begin{center}\rule{0.5\linewidth}{0.5pt}\end{center}

\hypertarget{app-primitive-influence-functions}{%
\subsection*{Primitive influence
functions}\label{app:primitive-influence-functions}}
\addcontentsline{toc}{subsection}{Primitive influence functions}

Define the matrix-valued influence terms

\[
\Psi_{Q,t}=U_tU_t'-Q,
\tag{A.2}
\]

\[
\Psi_{K,t}
=
\varepsilon_t\widetilde W_t'M_{XW}^{-1},
\tag{A.3}
\]

and the scalar

\[
\psi_{c,t}=\eta_t=B_t-c.
\tag{A.4}
\]

Let \(D_p\) be the duplication matrix and define

\[
\psi_t
=
\begin{pmatrix}
\operatorname{vech}(\Psi_{Q,t})\\
\operatorname{vec}(\Psi_{K,t})\\
\psi_{c,t}
\end{pmatrix}.
\tag{A.5}
\]

The joint long-run covariance is

\[
\Sigma
=
\sum_{\ell=-\infty}^{\infty}
E[\psi_t\psi_{t-\ell}'].
\tag{A.6}
\]

All off-diagonal blocks of \(\Sigma\) are retained. This matters because
\(R_t^L\) enters both \(U_t\) and the contrast \(Y_t=R_t^S-R_t^L\),
while balance-sheet capacity can comove with volatility and flow.

\hypertarget{app-lemma-1-asymptotic-linearity-of-the-primitives}{%
\subsection*{Lemma 1 --- asymptotic linearity of the
primitives}\label{app:lemma-1-asymptotic-linearity-of-the-primitives}}
\addcontentsline{toc}{subsection}{Lemma 1 --- asymptotic linearity
of the primitives}

\hypertarget{app-statement}{%
\subsubsection*{Statement}\label{app:statement}}
\addcontentsline{toc}{subsubsection}{Statement}

For each fixed \(P\in\mathcal P^S\), A1--A6 imply

\[
\sqrt T
\begin{pmatrix}
\operatorname{vech}(\widehat Q-Q)\\
\operatorname{vec}(\widehat K-K)\\
\widehat c-c
\end{pmatrix}
=
\frac1{\sqrt T}\sum_{t=1}^T\psi_t+o_P(1)
\Rightarrow N(0,\Sigma).
\tag{A.7}
\]

Moreover, for every triangular subclass
\(\mathcal P_T^S\subseteq\mathcal P^S\),

\[
\sup_{P\in\mathcal P_T^S}
P\left(
\left\|
\sqrt T(\widehat\vartheta-\vartheta(P))
-T^{-1/2}\sum_{t=1}^T\psi_t(P)
\right\|>\epsilon
\right)
\to0
\]

for every \(\epsilon>0\), and the normalized score sum obeys a uniform
multivariate central limit theorem.

Writing \(\Phi_{d_\vartheta}\) for standard Gaussian law and
\(\mathscr C\) for the convex Borel subsets of
\(\mathbb R^{d_\vartheta}\), the latter statement means

\[
\sup_{P\in\mathcal P_T^S}
\sup_{C\in\mathscr C}
\left|
P\left\{
\Sigma(P)^{-1/2}T^{-1/2}
\sum_{t=1}^T\psi_t(P)\in C
\right\}
-\Phi_{d_\vartheta}(C)
\right|
\to0.
\]

\hypertarget{app-proof}{%
\subsubsection*{Proof}\label{app:proof}}
\addcontentsline{toc}{subsubsection}{Proof}

For \(Q\), write \(\widehat U_t=U_t-(\widehat M_Q-M_Q)g_t\). Expanding
the sample residual covariance,

\[
\widehat Q-Q
=
T^{-1}\sum_{t=1}^T(U_tU_t'-Q)
-(\widehat M_Q-M_Q)
\left(T^{-1}\sum g_tU_t'\right)
\]

\[
\quad
-\left(T^{-1}\sum U_tg_t'\right)
(\widehat M_Q-M_Q)'
+O_P(\|\widehat M_Q-M_Q\|^2).
\tag{A.8}
\]

Because \(E[U_tg_t']=0\), the two cross-products are \(O_P(T^{-1})\),
and the last term is \(O_P(T^{-1})\). Therefore

\[
\sqrt T\,\operatorname{vech}(\widehat Q-Q)
=
T^{-1/2}\sum_{t=1}^T
\operatorname{vech}(U_tU_t'-Q)+o_P(1).
\tag{A.9}
\]

For \(K\), use \(\widehat K=\widehat M_{YW}\widehat M_{XW}^{-1}\),
\(M_{YW}=KM_{XW}\), and the inverse expansion:

\[
\widehat K-K
=
\{(\widehat M_{YW}-M_{YW})
-K(\widehat M_{XW}-M_{XW})\}
M_{XW}^{-1}
+o_P(T^{-1/2}).
\tag{A.10}
\]

To make the residualization step explicit, for any vectors \(A_t,D_t\)
let

\[
C_{AD}
=E[\widetilde A_t\widetilde D_t']
=E[A_tD_t']
-E[A_tg_t']E[g_tg_t']^{-1}E[g_tD_t'].
\]

The plug-in sample partial covariance has influence function

\[
\widetilde A_t\widetilde D_t'-C_{AD}.
\]

This follows by differentiating the three moment matrices in the second
display; the terms generated by the estimated projection coefficients
combine exactly into the product of population residuals. Applying the
identity to \((Y,W)\) and \((X,W)\), and using \(M_{YW}-KM_{XW}=0\),
gives the influence of their difference as

\[
(\widetilde Y_t-K\widetilde X_t)\widetilde W_t'
=\varepsilon_t\widetilde W_t'.
\]

Therefore

\[
\sqrt T(\widehat K-K)
=
\frac1{\sqrt T}\sum_{t=1}^T
\varepsilon_t\widetilde W_t'M_{XW}^{-1}
+o_P(1).
\tag{A.11}
\]

For \(c\), the sample-mean definition and \eqref{eq:observation-model} give

\[
\sqrt T(\widehat c-c)
=T^{-1/2}\sum_{t=1}^T\eta_t.
\tag{A.12}
\]

Stacking (A.9), (A.11), and (A.12), followed by the multivariate mixing
central limit theorem, proves (A.7) pointwise. For the uniform
statement, A2 and A4 bound all projection and inverse derivatives
uniformly. The sample moment errors are uniformly \(O_P(T^{-1/2})\), so
the quadratic Taylor remainders in (A.8) and (A.10) are uniformly
\(o_P(T^{-1/2})\). The common geometric bound in A1 and the common
\(2+\delta/2\) score moment implied by A2 permit the standard blocking
argument with a truncation level independent of \(P\); Cramér--Wold then
yields the uniform multivariate central limit theorem. \(\square\)

\hypertarget{app-audit-note}{%
\subsubsection*{Scope of linear residualization}\label{app:scope-linear-residualization}}
\addcontentsline{toc}{subsubsection}{Scope of linear residualization}

Equation (A.11) relies on linear residualization. If flexible or nonparametric estimation is
used for \(E[Y\mid Z]\), \(E[X\mid Z]\), or \(E[W\mid Z]\), the analysis
must replace this step by a cross-fitted Neyman-orthogonal score and
prove its remainder rate. The present results do not cover
that extension.

\hypertarget{app-primitive-to-operator-propagation}{%
\subsection*{Primitive-to-operator
propagation}\label{app:primitive-to-operator-propagation}}
\addcontentsline{toc}{subsection}{Primitive-to-operator propagation}

Let

\[
\widehat A=\widehat c\,\widehat Q^{-1}\widehat K.
\tag{A.13}
\]

\hypertarget{app-theorem-1}{%
\subsubsection*{Theorem 1}\label{app:theorem-1}}
\addcontentsline{toc}{subsubsection}{Theorem 1}

Under A1--A6, the map \(A(Q,K,c)=cQ^{-1}K\) is continuously Fréchet
differentiable on the parameter interior. For \(h=(H_Q,H_K,h_c)\),

\[
DA_\vartheta[h]
=
h_cQ^{-1}K
+cQ^{-1}H_K
-cQ^{-1}H_QQ^{-1}K.
\tag{A.14}
\]

If \(\vartheta=(\operatorname{vech}(Q)',\operatorname{vec}(K)',c)'\),
then

\[
\sqrt T\,\operatorname{vec}(\widehat A-A)
=
J_A(\vartheta)
\frac1{\sqrt T}\sum_{t=1}^T\psi_t
+o_P(1)
\Rightarrow
N(0,\Omega_A),
\tag{A.15}
\]

where

\[
J_A(\vartheta)
=
\left[
-c\{(Q^{-1}K)'\otimes Q^{-1}\}D_p,
\quad
c(I_p\otimes Q^{-1}),
\quad
\operatorname{vec}(Q^{-1}K)
\right],
\tag{A.16}
\]

\[
\Omega_A=J_A\Sigma J_A'.
\tag{A.17}
\]

Equivalently, the matrix influence function of \(\widehat A\) is

\[
\Psi_{A,t}
=
\psi_{c,t}Q^{-1}K
+cQ^{-1}\Psi_{K,t}
-cQ^{-1}\Psi_{Q,t}Q^{-1}K.
\tag{A.18}
\]

\hypertarget{app-proof-1}{%
\subsubsection*{Proof}\label{app:proof-1}}
\addcontentsline{toc}{subsubsection}{Proof}

Let \(\Delta_Q=\widehat Q-Q\), \(\Delta_K=\widehat K-K\), and
\(\Delta_c=\widehat c-c\). Since \(\lambda_{\min}(Q)\ge\underline q\)
and \(\|\Delta_Q\|=O_P(T^{-1/2})\), \(\widehat Q\) is invertible with
probability tending to one. The resolvent identity gives

\[
(Q+\Delta_Q)^{-1}
=Q^{-1}-Q^{-1}\Delta_QQ^{-1}
+R_Q,
\tag{A.19}
\]

where, on a neighborhood with \(\|Q^{-1}\Delta_Q\|<1\),

\[
\|R_Q\|
\le
\frac{\|Q^{-1}\|^3\|\Delta_Q\|^2}
{1-\|Q^{-1}\Delta_Q\|}
=O_P(T^{-1}).
\tag{A.20}
\]

Substitute (A.19) into

\[
\widehat A
=(c+\Delta_c)(Q+\Delta_Q)^{-1}(K+\Delta_K).
\]

The terms linear in the perturbations are

\[
\Delta_cQ^{-1}K
+cQ^{-1}\Delta_K
-cQ^{-1}\Delta_QQ^{-1}K.
\tag{A.21}
\]

Every remaining product contains either \(R_Q\) or at least two
primitive perturbations and is \(O_P(T^{-1})\). Thus (A.14) is the
Fréchet derivative and

\[
\sqrt T\,\operatorname{vec}(\widehat A-A)
=
J_A(\vartheta)
\sqrt T(\widehat\vartheta-\vartheta)
+o_P(1).
\]

Insert Lemma 1 and apply Slutsky's theorem to obtain (A.15)--(A.17).
Applying (A.14) directly to the matrix influence terms yields (A.18).
\(\square\)

\hypertarget{app-financial-decomposition}{%
\subsubsection*{Financial decomposition}\label{app:financial-decomposition}}
\addcontentsline{toc}{subsubsection}{Financial decomposition}

Equation (A.18) decomposes boundary-operator uncertainty into:

\begin{enumerate}
\def\labelenumi{\arabic{enumi}.}
\tightlist
\item
  capacity measurement error, \(\psi_{c,t}Q^{-1}K\);
\item
  cross-impact estimation error, \(cQ^{-1}\Psi_{K,t}\);
\item
  conditional-risk estimation error, \(-cQ^{-1}\Psi_{Q,t}Q^{-1}K\).
\end{enumerate}

Their covariances generally do not vanish. Reporting three separate
standard errors and adding them in quadrature would be incorrect unless
block independence is established.

\hypertarget{app-regular-active-spectrum}{%
\subsection*{Regular active
spectrum}\label{app:regular-active-spectrum}}
\addcontentsline{toc}{subsection}{Regular active spectrum}

Let \(\lambda_0\) attain
\(\min_{\lambda\in\operatorname{spec}(A)}\operatorname{Re}\lambda\). The regular set
\(\Theta_{\mathrm{reg}}\) contains parameters for which one of the
following holds:

\begin{enumerate}
\def\labelenumi{\arabic{enumi}.}
\tightlist
\item
  \(\lambda_0\in\mathbb R\) is algebraically simple and
\end{enumerate}

\[
\min_{\lambda\ne\lambda_0}
\{\operatorname{Re}\lambda-\operatorname{Re}\lambda_0\}
\ge\Delta_\lambda>0;
\tag{A.22}
\]

\begin{enumerate}
\def\labelenumi{\arabic{enumi}.}
\setcounter{enumi}{1}
\tightlist
\item
  \(\lambda_0\notin\mathbb R\), its conjugate \(\bar\lambda_0\) is the
  only other eigenvalue with the same real part, both are simple, and
  every remaining root satisfies (A.22).
\end{enumerate}

Let \(v_0,w_0\in\mathbb C^p\) be right and left eigenvectors:

\[
Av_0=\lambda_0v_0,
\qquad
w_0^*A=\lambda_0w_0^*,
\qquad
w_0^*v_0=1.
\tag{A.23}
\]

The normalization is possible for a simple eigenvalue. The eigenvalue
condition number is

\[
\kappa_\lambda=\frac{\|w_0\|\|v_0\|}{|w_0^*v_0|}.
\tag{A.24}
\]

Under (A.23), it reduces to \(\|w_0\|\|v_0\|\), but (A.24) is the
normalization-invariant definition.

\hypertarget{app-pointwise-inference-for-the-boundary-margin}{%
\subsection*{Pointwise inference for the boundary
margin}\label{app:pointwise-inference-for-the-boundary-margin}}
\addcontentsline{toc}{subsection}{Pointwise inference for the
boundary margin}

\hypertarget{app-theorem-2}{%
\subsubsection*{Theorem 2}\label{app:theorem-2}}
\addcontentsline{toc}{subsubsection}{Theorem 2}

Let \(\vartheta_0\in\Theta_{\mathrm{reg}}\), and suppose A1--A6 hold
pointwise at \(\vartheta_0\). Then the boundary margin is real
differentiable along real perturbations, with

\[
Dm_{\vartheta_0}[h]
=
\operatorname{Re}
\left\{
w_0^*DA_{\vartheta_0}[h]v_0
\right\}.
\tag{A.25}
\]

Define the scalar influence function

\[
\phi_{m,t}
=
\operatorname{Re}
\left\{
w_0^*\Psi_{A,t}v_0
\right\},
\tag{A.26}
\]

and its long-run variance

\[
V_m
=
\sum_{\ell=-\infty}^{\infty}
E[\phi_{m,t}\phi_{m,t-\ell}].
\tag{A.27}
\]

If \(0<V_m<\infty\), then

\[
\sqrt T\{m(\widehat\vartheta)-m(\vartheta_0)\}
=
\frac1{\sqrt T}\sum_{t=1}^T\phi_{m,t}
+o_P(1)
\Rightarrow N(0,V_m).
\tag{A.28}
\]

A feasible pointwise \(1-\alpha\) interval is

\[
I_T^\Delta
=
\left[
m(\widehat\vartheta)
-z_{1-\alpha/2}\sqrt{\widehat V_m/T},
\quad
m(\widehat\vartheta)
+z_{1-\alpha/2}\sqrt{\widehat V_m/T}
\right],
\tag{A.29}
\]

where \(\widehat V_m\) is a HAC estimate computed from the plug-in
analogue of (A.26).

\hypertarget{app-proof-2}{%
\subsubsection*{Proof}\label{app:proof-2}}
\addcontentsline{toc}{subsubsection}{Proof}

For a simple eigenvalue, analytic perturbation theory gives, for a real
matrix perturbation \(E\),

\[
\lambda(A+E)
=
\lambda_0+w_0^*Ev_0+o(\|E\|).
\tag{A.30}
\]

In the real-root case, the real-part separation in (A.22) ensures that
the perturbed continuation of \(\lambda_0\) remains the unique active
root in a neighborhood. Therefore

\[
m(A+E)-m(A)
=
\operatorname{Re}(w_0^*Ev_0)+o(\|E\|).
\tag{A.31}
\]

In the nonreal case, real perturbations preserve conjugacy. The two
active continuations have identical real parts, and (A.22) separates
them from all other roots. Equation (A.31) again holds when either
member of the pair is used.

Combine (A.31) with Theorem 1:

\[
\sqrt T\{m(\widehat\vartheta)-m(\vartheta_0)\}
=
\operatorname{Re}
\left\{
w_0^*\sqrt T(\widehat A-A)v_0
\right\}
+o_P(1).
\tag{A.32}
\]

Substituting (A.18) proves the asymptotic linear representation
(A.26)--(A.28). Consistency of the eigenvectors, primitive estimates and
HAC variance on the regular set, followed by Slutsky's theorem, yields
(A.29). \(\square\)

\hypertarget{app-nonnormality-as-an-inferential-amplifier}{%
\subsection*{Nonnormality as an inferential
amplifier}\label{app:nonnormality-as-an-inferential-amplifier}}
\addcontentsline{toc}{subsection}{Nonnormality as an inferential
amplifier}

From (A.25),

\[
|Dm_{\vartheta_0}[h]|
\le
\kappa_\lambda
\|DA_{\vartheta_0}[h]\|.
\tag{A.33}
\]

Two markets with the same primitive estimation error can therefore have
very different uncertainty about the regime. The difference is not
captured by \(\|\widehat A-A\|\) alone: it depends on the left-right
eigenvector geometry of the active mode.

The empirical reporting includes:

\begin{itemize}
\tightlist
\item
  the active eigenvalue or conjugate pair;
\item
  the real-part separation \(\widehat\Delta_\lambda\);
\item
  \(\widehat\kappa_\lambda\);
\item
  the three components of \(\widehat\phi_{m,t}\);
\item
  their full HAC covariance.
\end{itemize}

This makes the source of uncertainty economically interpretable instead
of hiding it inside a single threshold standard error.

\hypertarget{app-exact-scope-of-theorem-2}{%
\subsection*{Conditions for the regular approximation
}\label{app:exact-scope-of-theorem-2}}
\addcontentsline{toc}{subsection}{Conditions for the regular approximation}

The regular approximation is pointwise in \(\vartheta_0\), fixed-dimensional,
and conditional on strong IV relevance. It applies when the active spectrum is
a separated simple real root or a separated simple conjugate pair.

It is not uniformly valid as:

\[
\Delta_\lambda\downarrow0,
\qquad
\kappa_\lambda\uparrow\infty,
\qquad
\sigma_{\min}(M_{XW})\downarrow0,
\]

or when the parameter approaches a defective matrix. The first two
failures motivate the paper's spectral projection method; the third is a
separate weak-identification problem in the primitive estimator.

No data-dependent pretest selecting Theorem 2 whenever the estimated
spectrum ``looks regular'' will be treated as valid solely for that reason. The
projected procedure of Theorems 4--5 remains the primary classifier;
(A.29) is a transparent efficiency benchmark away from collisions.

\begin{center}\rule{0.5\linewidth}{0.5pt}\end{center}

\hypertarget{app-appendix-b.-nonregular-spectral-proofs}{%
\section*{Appendix B. Nonregular spectral
proofs}\label{app:appendix-b.-nonregular-spectral-proofs}}
\addcontentsline{toc}{section}{Appendix B. Nonregular spectral proofs}

\hypertarget{app-canonical-jordan-family}{%
\subsection*{Canonical Jordan
family}\label{app:canonical-jordan-family}}
\addcontentsline{toc}{subsection}{Canonical Jordan family}

Consider

\[
A(a,b)
=
\begin{pmatrix}
-\frac12+a&1\\
b&-\frac12+a
\end{pmatrix}.
\tag{B.1}
\]

Its eigenvalues are

\[
\lambda_\pm(a,b)
=-\frac12+a\pm\sqrt b,
\tag{B.2}
\]

where the principal square root is used. Consequently,

\[
m(a,b)
=
\begin{cases}
a-\sqrt b,&b\ge0,\\
a,&b<0.
\end{cases}
=a-\sqrt{b_+},
\qquad b_+=\max\{b,0\}.
\tag{B.3}
\]

The matrix \(A(0,0)\) has a size-two Jordan block at the equilibrium
boundary.

\hypertarget{app-proposition-3a-rate-change-at-the-jordan-boundary}{%
\subsubsection*{Proposition 3(a): rate change at the Jordan
boundary}\label{app:proposition-3a-rate-change-at-the-jordan-boundary}}
\addcontentsline{toc}{subsubsection}{Proposition 3(a): rate change at
the Jordan boundary}

Suppose a statistical submodel satisfies

\[
\widehat A_T
=
A(0,0)
+\frac{Z_T}{\sqrt T}e_2e_1',
\qquad
Z_T\Rightarrow Z,
\tag{B.4}
\]

where \(e_2e_1'\) has a one in position \((2,1)\), and \(Z\) has a
continuous nondegenerate distribution. Then

\[
T^{1/4}m(\widehat A_T)
\Rightarrow
-\sqrt{Z_+}.
\tag{B.5}
\]

If \(P(Z>0)>0\), the limit is non-Gaussian and has a point mass at zero
whenever \(P(Z\le0)>0\). In particular,

\[
\sqrt T\,m(\widehat A_T)
\]

is not tight.

\hypertarget{app-proof-3}{%
\subsubsection*{Proof}\label{app:proof-3}}
\addcontentsline{toc}{subsubsection}{Proof}

By (B.3), with \(a=0\) and \(b=Z_T/\sqrt T\),

\[
m(\widehat A_T)
=
-\sqrt{\left(\frac{Z_T}{\sqrt T}\right)_+}
=
-T^{-1/4}\sqrt{(Z_T)_+}.
\]

The map \(z\mapsto-\sqrt{z_+}\) is continuous. Equation (B.5) follows by
the continuous mapping theorem. On \(\{Z>0\}\), multiplying instead by
\(\sqrt T\) yields a term of order \(T^{1/4}\), so root-\(T\) tightness
fails. \(\square\)

\hypertarget{app-interpretation}{%
\subsubsection*{Interpretation}\label{app:interpretation}}
\addcontentsline{toc}{subsubsection}{Interpretation}

The primitive estimator remains regular and root-\(T\). The change of
rate is created entirely by the defective spectral map. This separates
Theorem 3 from weak identification of \(K\): the first-stage matrix may
be perfectly well conditioned while regime inference is nonregular.

\hypertarget{app-semisimple-ties-root-t-but-non-gaussian}{%
\subsection*{\texorpdfstring{Semisimple ties: root-\(T\) but
non-Gaussian}{5.3 Semisimple ties: root-T but non-Gaussian}}\label{app:semisimple-ties-root-t-but-non-gaussian}}
\addcontentsline{toc}{subsection}{Semisimple ties: root-\(T\) but
non-Gaussian}

Defectivity is not the only failure mode. Let

\[
A_0=-\frac12I_2.
\tag{B.6}
\]

Both active eigenvalues are semisimple. For every real \(2\times2\)
direction \(H\) and \(t>0\),

\[
m(A_0+tH)
=
t\min_{\lambda\in\operatorname{spec}(H)}
\operatorname{Re}\lambda.
\tag{B.7}
\]

\hypertarget{app-proposition-3b-nonlinear-directional-derivative}{%
\subsubsection*{Proposition 3(b): nonlinear directional
derivative}\label{app:proposition-3b-nonlinear-directional-derivative}}
\addcontentsline{toc}{subsubsection}{Proposition 3(b): nonlinear
directional derivative}

At \(A_0\), the margin is Hadamard directionally differentiable with

\[
m'_{A_0}(H)
=
\min_{\lambda\in\operatorname{spec}(H)}
\operatorname{Re}\lambda,
\tag{B.8}
\]

but it is not Fréchet differentiable.

If

\[
\sqrt T(\widehat A-A_0)\Rightarrow\mathbb Z,
\tag{B.9}
\]

then

\[
\sqrt T\,m(\widehat A)
\Rightarrow
\min_{\lambda\in\operatorname{spec}(\mathbb Z)}
\operatorname{Re}\lambda,
\tag{B.10}
\]

which is generally non-Gaussian.

\hypertarget{app-proof-4}{%
\subsubsection*{Proof}\label{app:proof-4}}
\addcontentsline{toc}{subsubsection}{Proof}

Because \(A_0\) is a scalar matrix,

\[
\operatorname{spec}(A_0+tH)
=
\left\{-\frac12+t\lambda:
\lambda\in\operatorname{spec}(H)\right\},
\]

which proves (B.7)--(B.8). The derivative is not linear. For example,
take

\[
H_1=
\begin{pmatrix}1&0\\0&0\end{pmatrix},
\qquad
H_2=
\begin{pmatrix}0&0\\0&1\end{pmatrix}.
\]

Then

\[
m'_{A_0}(H_1)=m'_{A_0}(H_2)=0,
\qquad
m'_{A_0}(H_1+H_2)=1.
\]

Hence no Fréchet derivative exists. Applying the directional delta
method to (B.9) gives (B.10). \(\square\)

\hypertarget{app-bootstrap-implication}{%
\subsubsection*{Bootstrap implication}\label{app:bootstrap-implication}}
\addcontentsline{toc}{subsubsection}{Bootstrap implication}

At a semisimple tie, an ordinary bootstrap that treats the margin as
smoothly differentiable cannot be presumed valid. The directional
derivative in (B.8) must itself be estimated, or inference must avoid
differentiating the margin. This paper takes the second route and
projects a confidence set for the primitive parameter.

\hypertarget{app-explicit-nonuniformity-of-the-simple-root-delta-interval}{%
\subsection*{Explicit nonuniformity of the simple-root delta
interval}\label{app:explicit-nonuniformity-of-the-simple-root-delta-interval}}
\addcontentsline{toc}{subsection}{Explicit nonuniformity of the
simple-root delta interval}

Proposition 3(a) works exactly at a defective root. The next theorem is
also covers a triangular sequence in which every true matrix has
two distinct real eigenvalues, so Theorem 2 would appear locally
applicable if its pointwise nature were ignored.

For fixed \(\kappa>0\), define

\[
b_T=\frac{\kappa^2}{\sqrt T},
\qquad
A_T=A(0,b_T).
\tag{B.11}
\]

The eigenvalues of \(A_T\) are distinct for every finite \(T\), with gap

\[
\Delta_{\lambda,T}
=2\kappa T^{-1/4},
\tag{B.12}
\]

and true margin

\[
m_T=-\kappa T^{-1/4}.
\tag{B.13}
\]

Assume the identified primitive coefficient obeys

\[
\widehat b_T
=b_T+\frac{Z_T}{\sqrt T}
+o_P(T^{-1/2}),
\qquad
Z_T\Rightarrow N(0,1),
\tag{B.14}
\]

while the remaining coordinates are known in this least-favourable
submodel or contribute only smaller-order terms. Then

\[
T^{1/4}\widehat m_T
\Rightarrow
-\sqrt{(\kappa^2+Z)_+}.
\tag{B.15}
\]

The simple-root derivative at the true \(b_T\) is

\[
\frac{\partial m}{\partial b}(0,b_T)
=-\frac{1}{2\sqrt{b_T}}
=-\frac{T^{1/4}}{2\kappa}.
\tag{B.16}
\]

Thus the oracle delta standard error is

\[
s_T^{\mathrm{or}}
=\frac{1}{2\kappa T^{1/4}}.
\tag{B.17}
\]

\hypertarget{app-theorem-3-failure-of-uniform-delta-coverage}{%
\subsubsection*{Theorem 3: failure of uniform delta
coverage}\label{app:theorem-3-failure-of-uniform-delta-coverage}}
\addcontentsline{toc}{subsubsection}{Theorem 3: failure of uniform delta
coverage}

Let

\[
I_T^{\mathrm{or}}
=
\left[
\widehat m_T
\pm z_{1-\alpha/2}s_T^{\mathrm{or}}
\right].
\tag{B.18}
\]

Then

\[
\lim_{T\to\infty}
P\{m_T\in I_T^{\mathrm{or}}\}
=
P\left\{
\left|
\kappa-\sqrt{(\kappa^2+Z)_+}
\right|
\le
\frac{z_{1-\alpha/2}}{2\kappa}
\right\}.
\tag{B.19}
\]

This limit is not generally \(1-\alpha\). For \(\alpha=0.05\),
\(\kappa=1\), and \(Z\sim N(0,1)\),

\[
\lim_{T\to\infty}
P\{m_T\in I_T^{\mathrm{or}}\}
=
\Phi(3.92)-\Phi(-1)
\approx0.840.
\tag{B.20}
\]

Therefore the pointwise delta interval is not uniformly valid on any
parameter class containing the sequence \(\{A_T\}\).

\hypertarget{app-proof-5}{%
\subsubsection*{Proof}\label{app:proof-5}}
\addcontentsline{toc}{subsubsection}{Proof}

Equations (B.3), (B.11), and (B.14) imply

\[
\widehat m_T
=
-T^{-1/4}
\sqrt{(\kappa^2+Z_T+o_P(1))_+}.
\]

Combining this expression with (B.13), multiplying the coverage
inequality by \(T^{1/4}\), and using the continuous mapping theorem
yields (B.19).

For \(\alpha=0.05\) and \(\kappa=1\), write

\[
q=\frac{z_{0.975}}2=0.98.
\]

Since \(1-q>0\), the coverage event is equivalent to

\[
(1-q)^2-1
\le Z\le
(1+q)^2-1,
\]

whose endpoints are \(-1\) and \(2.92\). This proves
(B.20). Because an oracle interval using the true local derivative
already misses nominal coverage, the pointwise theorem cannot supply a
uniform guarantee over this sequence. \(\square\)

\hypertarget{app-why-the-critical-scale-is-t-14}{%
\subsubsection*{\texorpdfstring{Why the critical scale is
\(T^{-1/4}\)}{Why the critical scale is T\^{}\{-1/4\}}}\label{app:why-the-critical-scale-is-t-14}}
\addcontentsline{toc}{subsubsection}{Why the critical scale is
\(T^{-1/4}\)}

For \(m(b)=-\sqrt b\), a Taylor expansion around \(b_T>0\) requires

\[
\frac{|\widehat b_T-b_T|}{b_T}=o_P(1).
\]

Since \(|\widehat b_T-b_T|=O_P(T^{-1/2})\), the requirement is

\[
b_T\sqrt T\to\infty,
\quad\text{or equivalently}\quad
\sqrt{b_T}\,T^{1/4}\to\infty.
\tag{B.21}
\]

The sequence (B.11) lies exactly at the transition where estimation
error and the squared eigenvalue gap have the same order. This is why
indexing experiments only by \(m_T=h/\sqrt T\) misses the defective
geometry.

\hypertarget{app-consequences-for-classification}{%
\subsection*{Consequences for
classification}\label{app:consequences-for-classification}}
\addcontentsline{toc}{subsection}{Consequences for classification}

Table \ref{tab:app-asymptotic-regimes} records the three inferential regions used in the proof discussion.

\begin{longtable}[]{@{}
  >{\raggedright\arraybackslash}p{(\columnwidth - 6\tabcolsep) * \real{0.231}}
  >{\raggedleft\arraybackslash}p{(\columnwidth - 6\tabcolsep) * \real{0.308}}
  >{\raggedright\arraybackslash}p{(\columnwidth - 6\tabcolsep) * \real{0.231}}
  >{\raggedright\arraybackslash}p{(\columnwidth - 6\tabcolsep) * \real{0.230}}@{}}
\caption{Local spectral geometry used in the nonregular proof discussion.}\label{tab:app-asymptotic-regimes}\\
\endfirsthead
\toprule\noalign{}
\begin{minipage}[b]{\linewidth}\raggedright
Spectral geometry
\end{minipage} & \begin{minipage}[b]{\linewidth}\raggedleft
Margin rate
\end{minipage} & \begin{minipage}[b]{\linewidth}\raggedright
First-order limit
\end{minipage} & \begin{minipage}[b]{\linewidth}\raggedright
Valid default
\end{minipage} \\
\midrule\noalign{}
\endhead
\bottomrule\noalign{}
\endlastfoot
Separated simple root/pair & \(T^{-1/2}\) & Gaussian & Theorem 2
pointwise benchmark \\
Semisimple active tie & \(T^{-1/2}\) & Nonlinear eigenvalue functional
of Gaussian matrix & Directional method or projection \\
Size-two Jordan limit & \(T^{-1/4}\) & Square-root transform with
possible atom & Primitive-set projection \\
\end{longtable}

The classifier must not choose one row using a non-uniform spectral
pretest. The primary procedure will instead construct a confidence
region for \((Q,K,c)\) and project the entire set through \(m\). Regular
geometry may be used to improve computation or report a benchmark, but
not to revoke the uniform guarantee.

\hypertarget{app-simulation-requirements-generated-by-theorem-3}{%
\subsection*{Implications for the simulation design
}\label{app:simulation-requirements-generated-by-theorem-3}}
\addcontentsline{toc}{subsection}{Implications for the simulation design}

The structural Monte Carlo must include the geometries that distinguish
the regular and nonregular mechanisms:

\begin{enumerate}
\def\labelenumi{\arabic{enumi}.}
\tightlist
\item
  a symmetric separated spectrum;
\item
  a nonnormal but diagonalizable spectrum;
\item
  the near-Jordan transition gap \(2\kappa T^{-1/4}\);
\item
  an exact Jordan block;
\item
  a semisimple repeated root;
\item
  a complex conjugate active pair.
\end{enumerate}

Each geometry is crossed with fixed positive and negative margins, local
margins of order \(T^{-1/2}\), and the exact boundary. The feasible
complete-design comparison reports:

\begin{itemize}
\tightlist
\item
  the plug-in classifier;
\item
  the feasible delta interval;
\item
  a deterministic norm envelope;
\item
  the projected primitive confidence set.
\end{itemize}

The oracle delta interval is the analytical counterexample in Theorem 3,
where it isolates approximation failure from variance-estimation
failure. It is not a feasible Monte Carlo method and is not relabelled
as one. A directional bootstrap would require a separately specified
active-cluster estimator and resampling law; no such procedure is
proposed or implemented in this paper, so it is not listed as an
unexecuted complete-design requirement.

\hypertarget{app-appendix-c.-uniform-projection-and-certified-classification}{%
\section*{Appendix C. Uniform projection and certified
classification}\label{app:appendix-c.-uniform-projection-and-certified-classification}}
\addcontentsline{toc}{section}{Appendix C. Uniform projection and
certified classification}

\hypertarget{app-uniform-primitive-confidence-region}{%
\subsection*{Uniform primitive confidence
region}\label{app:uniform-primitive-confidence-region}}
\addcontentsline{toc}{subsection}{Uniform primitive confidence
region}

Let \(\mathcal P_T^S\subseteq\mathcal P^S\) be any triangular subclass
governed by the common constants in A1--A6. To make later theorem
statements compact, write:

\begin{itemize}
\tightlist
\item
  \textbf{U1} for the uniform dependence, moment and fixed-design
  conditions A1--A3;
\item
  \textbf{U2} for strong IV relevance and primitive interiority in
  A4--A5, together with the long-run covariance eigenvalue bounds in A6;
\item
  \textbf{U3} for the fixed kernel and bandwidth construction of
  \(\widehat\Sigma\) in A6.
\end{itemize}

These labels now refer only to primitive sampling and design
restrictions. They do not assume the Wald conclusion.

\hypertarget{app-lemma-4a-primitive-conditions-imply-the-uniform-wald-approximation}{%
\subsubsection*{Lemma 4(a): primitive conditions imply the uniform Wald
approximation}\label{app:lemma-4a-primitive-conditions-imply-the-uniform-wald-approximation}}
\addcontentsline{toc}{subsubsection}{Lemma 4(a): primitive conditions
imply the uniform Wald approximation}

Under U1--U3,

\[
\sup_{P\in\mathcal P_T^S}
P\{\|\widehat\Sigma-\Sigma(P)\|>\epsilon\}
\to0
\qquad(\epsilon>0),
\tag{C.1}
\]

and

\[
\sup_{P\in\mathcal P_T^S}
\sup_{x\in\mathbb R}
\left|
P\left\{
T(\widehat\vartheta-\vartheta(P))'
\widehat\Sigma^{-1}
(\widehat\vartheta-\vartheta(P))
\le x
\right\}
-F_{\chi^2_{d_\vartheta}}(x)
\right|
\to0.
\tag{C.2}
\]

\hypertarget{app-proof-6}{%
\paragraph*{Proof}\label{app:proof-6}}
\addcontentsline{toc}{paragraph}{Proof}

We give the uniform steps because pointwise Cram\'er--Wold convergence
alone would not establish the convex-set approximation used by the Wald
statistic. Write

\[
S_T(P)=T^{-1/2}\sum_{t=1}^T\psi_t(P).
\]

Choose big-block length \(\ell_T=\lfloor T^{1/3}\rfloor\) and separating
gap length \(g_T=\lceil C\log T\rceil\), where \(C\) is large enough that

\[
\frac{T}{\ell_T}\sup_{P\in\mathcal P_T^S}\beta_P(g_T)\longrightarrow0.
\tag{C.3}
\]

The total variance of the discarded gaps is \(O(g_T/\ell_T)=o(1)\)
uniformly, by the covariance inequality for absolutely regular
sequences and the common score moment bound. Berbee coupling then
replaces the retained big blocks by independent copies with probability
error bounded by the left side of (C.3). For some \(\eta>0\), A2 and the
geometric mixing bound imply the uniform block-moment inequality

\[
\sup_P E_P\left\|\sum_{t=1}^{\ell_T}\psi_t(P)\right\|^{2+\eta}
\le C_\eta\ell_T^{1+\eta/2}.
\tag{C.4}
\]

Consequently the Lyapunov ratio of the independent block array is
bounded by a constant times

\[
\left(\frac{\ell_T}{T}\right)^{\eta/2}\longrightarrow0.
\tag{C.5}
\]

The fixed-dimensional convex-set normal approximation for independent
vectors (Bentkus, 1986), followed by the coupling and gap bounds, yields

\[
\sup_{P\in\mathcal P_T^S}\sup_{C\in\mathscr C}
\left|
P\{\Sigma(P)^{-1/2}S_T(P)\in C\}-\Phi_{d_\vartheta}(C)
\right|\longrightarrow0.
\tag{C.6}
\]

The eigenvalue bounds in A6 make the standardization uniform. Lemma 1
then transfers (C.6) from the score sum to
\(\sqrt T(\widehat\vartheta-\vartheta(P))\), since enlargements of convex
sets by an \(o_P(1)\) radius have uniformly vanishing Gaussian boundary
probability in fixed dimension.

It remains to studentize. Geometric absolute regularity and (C.4) imply
a uniformly summable covariance envelope. Splitting the population
series at \(b_T\), the omitted tail is therefore \(o(1)\) uniformly.
Continuity of \(k\) at every fixed lag and \(b_T\to\infty\) remove the
kernel bias. The retained oracle autocovariances obey

\[
\sup_P\|\widehat\Sigma^{\,o}-E_P\widehat\Sigma^{\,o}\|
=O_P\{(b_T/T)^{1/2}+b_T/T\}=o_P(1).
\tag{C.7}
\]

The plug-in scores are smooth functions of the same finite-dimensional
sample moments used in Lemma 1. A2 and A4 give, uniformly over retained
lags,

\[
\max_{0\le j\le b_T}
\|\widehat\Gamma_j-\widehat\Gamma_j^{\,o}\|=O_P(T^{-1/2}),
\]

so their accumulated contribution is
\(O_P(b_TT^{-1/2})=o_P(1)\). These bounds give (C.1); they are the
fixed-dimensional geometrically mixing specialization of Andrews
(1991) and de Jong and Davidson (2000).

Finally, the common lower eigenvalue bound makes covariance inversion
uniformly continuous. Combining (C.6), Lemma 1 and (C.1), and using the
continuity of the chi-square distribution, gives (C.2). \(\square\)

For emphasis, the class on which this conclusion holds retains the fixed
first-stage separation

\[
\inf_{P\in\mathcal P_T^S}
\sigma_{\min}\{M_{XW}(P)\}
\ge\underline\kappa_W>0.
\tag{C.8}
\]

Let

\[
q_{d_\vartheta,1-\alpha}
=F^{-1}_{\chi^2_{d_\vartheta}}(1-\alpha),
\]

and define the constrained Wald region

\[
\mathcal C_T(1-\alpha)
=
\left\{
\vartheta\in\Theta:
T(\widehat\vartheta-\vartheta)'
\widehat\Sigma^{-1}
(\widehat\vartheta-\vartheta)
\le q_{d_\vartheta,1-\alpha}
\right\}.
\tag{C.9}
\]

If \(\widehat\Sigma\) is not positive definite, define
\(\mathcal C_T=\Theta\) and report a primitive-covariance failure rather
than inserting an undeclared ridge. Under U1--U3 this event has
probability tending uniformly to zero by (C.1) and the eigenvalue lower
bound in A6.

Intersecting with \(\Theta\) enforces \(Q\succ0\), \(c>0\), fixed units
and the compactness bounds in (A.1). It does not reduce coverage
because the true parameter belongs to \(\Theta\).

\hypertarget{app-lemma-4b-uniform-primitive-coverage}{%
\subsubsection*{Lemma 4(b): uniform primitive
coverage}\label{app:lemma-4b-uniform-primitive-coverage}}
\addcontentsline{toc}{subsubsection}{Lemma 4(b): uniform primitive
coverage}

Under U1--U3,

\[
\liminf_{T\to\infty}
\inf_{P\in\mathcal P_T^S}
P\{\vartheta(P)\in\mathcal C_T(1-\alpha)\}
\ge1-\alpha.
\tag{C.10}
\]

\hypertarget{app-proof-7}{%
\subsubsection*{Proof}\label{app:proof-7}}
\addcontentsline{toc}{subsubsection}{Proof}

On the event that \(\widehat\Sigma\) is positive definite, membership of
the true \(\vartheta(P)\) in (C.9) is exactly the event that the quadratic
form in (C.2) does not exceed \(q_{d_\vartheta,1-\alpha}\). If
\(\widehat\Sigma\) is not positive definite, membership holds by the
conservative definition \(\mathcal C_T=\Theta\). Uniform convergence to
the continuous chi-square distribution therefore gives (C.10).
Intersecting the Wald ellipsoid with \(\Theta\) otherwise leaves the
true-parameter event unchanged because \(\vartheta(P)\in\Theta\).
\(\square\)

\hypertarget{app-identification-boundary}{%
\subsubsection*{Identification boundary}\label{app:identification-boundary}}
\addcontentsline{toc}{subsubsection}{Identification boundary}

The spectral boundary \(m=0\) is allowed inside \(\mathcal P_T^S\).
Drifting or weak instruments, singular \(Q\), an unidentified capacity
scale, growing dimension and a data-selected state dictionary are not.
On a sequence with \(\kappa_T\to0\), the uniform inverse bound used in
Lemma 1 fails; if \(\sqrt T\kappa_T=O(1)\), even the linear IV expansion
fails. In either case (C.9) is not justified. Theorem 4 remains a
deterministic projection statement for any valid primitive confidence
set, but this paper proves its probability guarantee only for the
strongly identified Wald front end.

\hypertarget{app-exact-projected-margin-set}{%
\subsection*{Projected margin
set}\label{app:exact-projected-margin-set}}
\addcontentsline{toc}{subsection}{Projected margin set}

Define

\[
L_T
=
\inf_{\vartheta\in\mathcal C_T}m(\vartheta),
\qquad
U_T
=
\sup_{\vartheta\in\mathcal C_T}m(\vartheta).
\tag{C.11}
\]

If \(\mathcal C_T=\varnothing\), set \(L_T=-\infty\), \(U_T=+\infty\)
and report a primitive-estimation failure. Otherwise, compactness of
\(\Theta\), closedness of \(\mathcal C_T\), and continuity of the
unordered spectrum imply that both extrema are attained.

\hypertarget{app-theorem-4a-uniformly-valid-projected-interval}{%
\subsubsection*{Theorem 4(a): uniformly valid projected
interval}\label{app:theorem-4a-uniformly-valid-projected-interval}}
\addcontentsline{toc}{subsubsection}{Theorem 4(a): uniformly valid
projected interval}

Under U1--U3,

\[
\liminf_{T\to\infty}
\inf_{P\in\mathcal P_T^S}
P\left\{
m(\vartheta(P))\in[L_T,U_T]
\right\}
\ge1-\alpha.
\tag{C.12}
\]

This result holds at simple roots, semisimple ties, complex active pairs
and defective matrices.

\hypertarget{app-proof-8}{%
\subsubsection*{Proof}\label{app:proof-8}}
\addcontentsline{toc}{subsubsection}{Proof}

On the event \(\vartheta(P)\in\mathcal C_T\), the definitions in (C.11)
imply

\[
L_T\le m(\vartheta(P))\le U_T.
\]

Therefore

\[
P\{m(\vartheta(P))\in[L_T,U_T]\}
\ge
P\{\vartheta(P)\in\mathcal C_T\}.
\]

Apply Lemma 4 and take the uniform lower limit. No spectral derivative
is used. \(\square\)

\hypertarget{app-certified-outer-bounds}{%
\subsection*{Certified outer bounds}\label{app:certified-outer-bounds}}
\addcontentsline{toc}{subsection}{Certified outer bounds}

The exact extrema in (C.11) are generally nonconvex and nonsmooth. A
floating-point optimizer that returns a candidate minimum or maximum
does not establish either one. The implementation must return outward
bounds

\[
\underline L_T\le L_T,
\qquad
\overline U_T\ge U_T.
\tag{C.13}
\]

Then

\[
[L_T,U_T]\subseteq
[\underline L_T,\overline U_T].
\]

\hypertarget{app-certificate-contract}{%
\subsubsection*{Certificate requirements}\label{app:certificate-contract}}
\addcontentsline{toc}{subsubsection}{Certificate requirements}

A numerical routine is theorem-compatible only if, for every retained
parameter box \(\mathbb B\), it produces an interval

\[
\mathfrak M(\mathbb B)
=
[\ell(\mathbb B),u(\mathbb B)]
\tag{C.14}
\]

satisfying

\[
\{m(\vartheta):\vartheta\in\mathbb B\cap\mathcal C_T\}
\subseteq
\mathfrak M(\mathbb B).
\tag{C.15}
\]

If the routine cannot verify (C.15), it must return a wider rigorous
enclosure, including the global fallback

\[
\left[
\frac12-\frac{\bar c\bar k}{\underline q},
\frac12+\frac{\bar c\bar k}{\underline q}
\right],
\tag{C.16}
\]

which follows from

\[
|\lambda\{cQ^{-1}K\}|
\le\|cQ^{-1}K\|
\le\frac{\bar c\bar k}{\underline q}.
\]

Failure to obtain a sharp enclosure may force an unresolved decision,
but cannot create a wrong resolved declaration.

\hypertarget{app-algorithm-1-verified-spectral-branch-and-bound}{%
\subsection*{Algorithm 1 --- verified spectral branch and
bound}\label{app:algorithm-1-verified-spectral-branch-and-bound}}
\addcontentsline{toc}{subsection}{Algorithm 1 --- verified spectral
branch and bound}

Write the ellipsoid in whitened coordinates:

\[
\vartheta(z)
=
\widehat\vartheta
+T^{-1/2}\widehat\Sigma^{1/2}z,
\qquad
\|z\|_2\le
\sqrt{q_{d_\vartheta,1-\alpha}},
\tag{C.17}
\]

and intersect its image with \(\Theta\).

\textbf{Input:} \(\widehat\vartheta,\widehat\Sigma,\Theta,\alpha\),
numerical tolerance \(\epsilon_T^{\mathrm{num}}\).\\
\textbf{Output:} \(\underline L_T,\overline U_T\), feasible inner values
\(L_T^{\mathrm{in}},U_T^{\mathrm{in}}\), and a certificate log.

\begin{enumerate}
\def\labelenumi{\arabic{enumi}.}
\tightlist
\item
  Cover the whitened ball in (C.17) by outward-rounded boxes.
\item
  Discard a box only after certifying that it does not intersect the
  ball or \(\Theta\).
\item
  Map every surviving box through \(Q,K,c\) using interval arithmetic.
\item
  Enclose \(Q^{-1}K\) with a validated interval linear solver; never
  invert an interval matrix entrywise.
\item
  Enclose every eigenvalue cluster of \(cQ^{-1}K\), including its
  multiplicity, using a verified eigensolver or interval
  characteristic-root isolation.
\item
  Convert the certified cluster rectangles into \(\ell(\mathbb B)\) and
  \(u(\mathbb B)\) for the minimum real part plus \(1/2\).
\item
  Evaluate feasible points to obtain
\end{enumerate}

\[
L_T\le L_T^{\mathrm{in}},
\qquad
U_T^{\mathrm{in}}\le U_T.
\tag{C.18}
\]

\begin{enumerate}
\def\labelenumi{\arabic{enumi}.}
\setcounter{enumi}{7}
\tightlist
\item
  Bisect the box contributing the largest unresolved global gap.
\item
  Stop when a regime sign is certified, the requested tolerance is
  reached, or the computational budget is exhausted.
\end{enumerate}

At any stopping time define

\[
\underline L_T=\min_{\mathbb B}\ell(\mathbb B),
\qquad
\overline U_T=\max_{\mathbb B}u(\mathbb B).
\tag{C.19}
\]

The certificate log records outward-rounding mode, boxes discarded,
cluster counts, fallback calls, inner candidates and final gaps.

\hypertarget{app-numerical-convergence-condition}{%
\subsubsection*{Numerical convergence
condition}\label{app:numerical-convergence-condition}}
\addcontentsline{toc}{subsubsection}{Numerical convergence condition}

For resolution consistency, require

\[
0\le
L_T^{\mathrm{in}}-\underline L_T
\le\epsilon_T^{\mathrm{num}},
\qquad
0\le
\overline U_T-U_T^{\mathrm{in}}
\le\epsilon_T^{\mathrm{num}},
\qquad
\epsilon_T^{\mathrm{num}}=o_P(1).
\tag{C.20}
\]

The error-control theorem below requires only the outward relations
(C.13), not convergence in (C.20).

\hypertarget{app-implementation-scope}{%
\subsubsection*{Dimension and certification guarantees}\label{app:implementation-scope}}
\addcontentsline{toc}{subsubsection}{Dimension and certification guarantees}

The baseline experiments use small fixed \(p\), where verified
interval computation is feasible. For larger \(p\), a conservative
pseudospectral or norm envelope may replace sharp branch and bound. It
may increase the unresolved rate but must preserve (C.13).

\hypertarget{app-reference-implementation-record}{%
\subsubsection*{Two-dimensional certification
}\label{app:reference-implementation-record}}
\addcontentsline{toc}{subsubsection}{Two-dimensional certification}

For \(p=2\), Algorithm 1 maps the affine Wald region through
outward-rounded interval arithmetic, uses a validated analytic solve for the
symmetric \(2\times2\) matrix \(Q\), and isolates the two characteristic roots
through the discriminant
identity

\[
\Delta(A)=(a_{11}-a_{22})^2+4a_{12}a_{21}.
\tag{C.21}
\]

This form remains valid at complex pairs, semisimple collisions and defective
Jordan points. If a box cannot certify positive invertibility of its entire
interval image, the global envelope in (C.16) is used. For \(p>2\), only that
global envelope is asserted here; an ordinary eigendecomposition is not a
verified substitute.

The Wald front end returns an outward radius \(\bar r_\alpha\)
satisfying

\[
P\{\chi^2_{d_\vartheta}>\bar r_\alpha^2\}\le\alpha
\tag{C.22}
\]

and an affine map \(M_T\) for which interval Cholesky certifies

\[
T M_TM_T'-\widehat\Sigma\succeq0.
\tag{C.23}
\]

Positive-semidefinite ordering implies that the nominal Wald ellipsoid
is contained in \(\{\widehat\vartheta+M_Tz:\|z\|_2\le\bar r_\alpha\}\).
For the \(p=2\) baseline, \(d_\vartheta=8\), and the even-degree
chi-square survival function is evaluated by its finite series with
directed high-precision decimal rounding. A Cantelli moment bound is
available as a more conservative fallback. The covariance factor is
inflated only until (C.23) is verified; if scaled Cholesky fails, an
isotropic Gershgorin envelope is attempted.

Numerical validation covers exact diagonal roots, complex pairs, semisimple
collisions, a defective Jordan boundary, interval containment, covariance
dominance, chi-square tail control, rejection of invalid covariance inputs,
the higher-dimensional fallback, and the capacity buffer under overshoot.
These checks validate the stated two-dimensional calculation, not every
computing environment. U1--U3 remain primitive statistical restrictions:
numerical certification cannot rescue an inconsistent covariance estimator or
an invalid identification design.

\hypertarget{app-theorem-4b-validity-with-numerical-certificates}{%
\subsection*{Theorem 4(b): validity with numerical
certificates}\label{app:theorem-4b-validity-with-numerical-certificates}}
\addcontentsline{toc}{subsection}{Theorem 4(b): validity with
numerical certificates}

If (C.13) holds almost surely for every numerical output, then

\[
\liminf_{T\to\infty}
\inf_{P\in\mathcal P_T^S}
P\left\{
m(\vartheta(P))
\in[\underline L_T,\overline U_T]
\right\}
\ge1-\alpha.
\tag{C.24}
\]

\hypertarget{app-proof-9}{%
\subsubsection*{Proof}\label{app:proof-9}}
\addcontentsline{toc}{subsubsection}{Proof}

The interval in (C.24) contains \([L_T,U_T]\) by (C.13). Apply Theorem
4(a). \(\square\)

\hypertarget{app-three-way-regime-classifier}{%
\subsection*{Three-way regime
classifier}\label{app:three-way-regime-classifier}}
\addcontentsline{toc}{subsection}{Three-way regime classifier}

Define the spectral classifier

\[
\delta_T
=
\begin{cases}
\mathrm{SUB},&\underline L_T>0,\\
\mathrm{SUP}_{\mathrm{spec}},&\overline U_T<0,\\
\mathrm{UNR},&\underline L_T\le0\le\overline U_T.
\end{cases}
\tag{C.25}
\]

Here \(\mathrm{UNR}\) means statistically or computationally unresolved.
It includes both genuine near-boundary uncertainty and failure of the
verified solver to sharpen its fallback bounds.

Let \(\mathcal G\subseteq\Theta\) denote the set satisfying all
additional regularity conditions required for the economic supercritical
construction. The stronger label \(\mathrm{SUP}_{\mathrm{econ}}\) is
permitted only if

\[
\overline U_T<0
\quad\text{and}\quad
\mathcal C_T\subseteq\mathcal G
\tag{C.26}
\]

are both certified. Otherwise the correct resolved statement is spectral
supercriticality only.

\hypertarget{app-uniform-control-of-wrong-resolved-declarations}{%
\subsection*{Uniform control of wrong resolved
declarations}\label{app:uniform-control-of-wrong-resolved-declarations}}
\addcontentsline{toc}{subsection}{Uniform control of wrong resolved
declarations}

Define the wrong-resolved event

\[
\mathcal E_T
=
\{\delta_T=\mathrm{SUB},\ m(\vartheta(P))\le0\}
\cup
\{\delta_T=\mathrm{SUP}_{\mathrm{spec}},\ m(\vartheta(P))\ge0\}.
\tag{C.27}
\]

\hypertarget{app-theorem-5a}{%
\subsubsection*{Theorem 5(a)}\label{app:theorem-5a}}
\addcontentsline{toc}{subsubsection}{Theorem 5(a)}

Under U1--U3 and the certificate relation (C.13),

\[
\limsup_{T\to\infty}
\sup_{P\in\mathcal P_T^S}
P(\mathcal E_T)
\le\alpha.
\tag{C.28}
\]

\hypertarget{app-proof-10}{%
\subsubsection*{Proof}\label{app:proof-10}}
\addcontentsline{toc}{subsubsection}{Proof}

If \(\delta_T=\mathrm{SUB}\), then \(\underline L_T>0\). A true margin
\(m(\vartheta(P))\le0\) lies outside \([\underline L_T,\overline U_T]\).
The same reasoning applies when
\(\delta_T=\mathrm{SUP}_{\mathrm{spec}}\) but \(m(\vartheta(P))\ge0\).
Hence

\[
\mathcal E_T
\subseteq
\{m(\vartheta(P))\notin
[\underline L_T,\overline U_T]\}.
\]

Apply (C.24). \(\square\)

The theorem controls wrong declarations, not the unresolved probability.
A method can satisfy (C.28) trivially by always returning
\(\mathrm{UNR}\); resolution is addressed separately.

\hypertarget{app-resolution-consistency-away-from-the-boundary}{%
\subsection*{Resolution consistency away from the
boundary}\label{app:resolution-consistency-away-from-the-boundary}}
\addcontentsline{toc}{subsection}{Resolution consistency away from
the boundary}

Assume the confidence region concentrates uniformly:

\[
\forall\eta>0:\quad
\sup_{P\in\mathcal P_T^S}
P\left\{
\sup_{\vartheta\in\mathcal C_T}
\|\vartheta-\vartheta(P)\|>\eta
\right\}
\to0.
\tag{C.29}
\]

For the Wald ellipsoid, (C.29) follows from uniform consistency of
\(\widehat\vartheta\), bounded eigenvalues of \(\widehat\Sigma\), fixed
\(d_\vartheta\), and the compact parameter restrictions.

Let

\[
\mathcal P_T^S(\varepsilon)
=
\{P\in\mathcal P_T^S:
|m(\vartheta(P))|\ge\varepsilon\},
\qquad\varepsilon>0.
\tag{C.30}
\]

\hypertarget{app-theorem-5b}{%
\subsubsection*{Theorem 5(b)}\label{app:theorem-5b}}
\addcontentsline{toc}{subsubsection}{Theorem 5(b)}

Under (C.13), (C.20), and (C.29),

\[
\inf_{\substack{P\in\mathcal P_T^S(\varepsilon)\\m(\vartheta(P))>0}}
P\{\delta_T=\mathrm{SUB}\}
\to1,
\tag{C.31}
\]

\[
\inf_{\substack{P\in\mathcal P_T^S(\varepsilon)\\m(\vartheta(P))<0}}
P\{\delta_T=\mathrm{SUP}_{\mathrm{spec}}\}
\to1.
\tag{C.32}
\]

\hypertarget{app-proof-11}{%
\subsubsection*{Proof}\label{app:proof-11}}
\addcontentsline{toc}{subsubsection}{Proof}

The map \(\vartheta\mapsto A(\vartheta)\) is continuous on compact \(\Theta\),
and the unordered eigenvalue set is continuous in the matrix entries.
Therefore \(m\) is uniformly continuous on \(\Theta\). Equation (C.29)
implies

\[
\sup_{\vartheta\in\mathcal C_T}
|m(\vartheta)-m(\vartheta(P))|
=o_P(1)
\tag{C.33}
\]

uniformly. Thus, on the subcritical separated class,
\(L_T>\varepsilon/2\) with probability tending uniformly to one; on the
supercritical separated class, \(U_T<-\varepsilon/2\) with probability
tending uniformly to one.

The inner/outer gaps in (C.20) imply

\[
L_T-\underline L_T\le\epsilon_T^{\mathrm{num}},
\qquad
\overline U_T-U_T\le\epsilon_T^{\mathrm{num}}.
\]

Since \(\epsilon_T^{\mathrm{num}}=o_P(1)\), the certified bounds inherit
the corresponding sign with probability tending to one. Equations
(C.31)--(C.32) follow. \(\square\)

No eigenvalue gap or diagonalizability condition is needed for
consistency at a fixed positive distance from the boundary.

\hypertarget{app-why-forced-binary-classification-cannot-resolve-the-local-band}{%
\subsection*{Local nonresolution and binary classification
}\label{app:why-forced-binary-classification-cannot-resolve-the-local-band}}
\addcontentsline{toc}{subsection}{Local nonresolution and binary classification}

The value of abstention can be shown in a one-dimensional Gaussian
subexperiment embedded in the primitive model. Set

\[
Q=1,\qquad c=1,\qquad K=-\frac12+a,
\]

so that \(m(a)=a\). Suppose

\[
\widehat a\sim N(a,T^{-1})
\]

and consider

\[
a_T^+=\frac h{\sqrt T},
\qquad
a_T^-=-\frac h{\sqrt T},
\qquad h>0.
\tag{C.34}
\]

\hypertarget{app-theorem-5c}{%
\subsubsection*{Theorem 5(c)}\label{app:theorem-5c}}
\addcontentsline{toc}{subsubsection}{Theorem 5(c)}

For any forced binary classifier

\[
\delta_T^{\mathrm{bin}}\in
\{\mathrm{SUB},\mathrm{SUP}_{\mathrm{spec}}\},
\]

\[
P_{a_T^+}
\{\delta_T^{\mathrm{bin}}=\mathrm{SUP}_{\mathrm{spec}}\}
+
P_{a_T^-}
\{\delta_T^{\mathrm{bin}}=\mathrm{SUB}\}
\ge
2\{1-\Phi(h)\}.
\tag{C.35}
\]

Consequently,

\[
\max\left[
P_{a_T^+}
\{\text{wrong}\},
P_{a_T^-}
\{\text{wrong}\}
\right]
\ge1-\Phi(h)>0.
\tag{C.36}
\]

No forced binary classifier is uniformly consistent over a root-\(T\)
neighborhood of the boundary.

\hypertarget{app-proof-12}{%
\subsubsection*{Proof}\label{app:proof-12}}
\addcontentsline{toc}{subsubsection}{Proof}

Under \(a_T^\pm\), the sufficient statistic \(\sqrt T\,\widehat a\) has
laws \(N(\pm h,1)\). For testing two simple hypotheses, the sum of
type-I and type-II errors of any decision rule is bounded below by one
minus the total-variation distance. For two unit-variance normals with
means \(\pm h\),

\[
\operatorname{TV}\{N(h,1),N(-h,1)\}
=2\Phi(h)-1.
\]

Substitution gives (C.35), and (C.36) follows because the maximum is at
least half the sum. \(\square\)

The three-way procedure does not eliminate this information bound. It
converts unavoidable local ambiguity into \(\mathrm{UNR}\) rather than a
statistically unjustified sign.

\hypertarget{app-error-resolution-and-abstention}{%
\subsection*{Error, resolution and
abstention}\label{app:error-resolution-and-abstention}}
\addcontentsline{toc}{subsection}{Error, resolution and abstention}

The three performance measures must be reported separately:

\[
\mathrm{WrongResolved}
=P(\mathcal E_T),
\tag{C.37}
\]

\[
\mathrm{Resolution}
=P(\delta_T\ne\mathrm{UNR}),
\tag{C.38}
\]

\[
\mathrm{CorrectResolution}
=P\{\delta_T
\text{ is resolved and correct}\}.
\tag{C.39}
\]

A method can reduce wrong classifications by abstaining more often.
Economic evaluation must therefore add the cost of delay or conservative
deployment rather than ranking procedures by error alone.

\hypertarget{app-appendix-d.-capacity-guarantees}{%
\section*{Appendix D. Capacity
guarantees}\label{app:appendix-d.-capacity-guarantees}}
\addcontentsline{toc}{section}{Appendix D. Capacity guarantees}

\hypertarget{app-destabilizing-intensity}{%
\subsection*{Destabilizing
intensity}\label{app:destabilizing-intensity}}
\addcontentsline{toc}{subsection}{Destabilizing intensity}

For a future capacity decision, separate the operator into

\[
B(Q,K)=Q^{-1}K
\tag{D.1}
\]

and define its destabilizing intensity

\[
d(Q,K)
=
\max\left\{
0,
-\min_{\lambda\in\operatorname{spec}(B(Q,K))}
\operatorname{Re}\lambda
\right\}.
\tag{D.2}
\]

For any deployed capacity \(a\ge0\),

\[
m(Q,K,a)
=
\frac12
+a\min_{\lambda\in\operatorname{spec}(B)}
\operatorname{Re}\lambda.
\tag{D.3}
\]

Hence

\[
m(Q,K,a)
=
\begin{cases}
\frac12-ad(Q,K),&
\min\operatorname{Re}\lambda(B)<0,\\
\frac12+a\min\operatorname{Re}\lambda(B)
\ge\frac12,&
\min\operatorname{Re}\lambda(B)\ge0.
\end{cases}
\tag{D.4}
\]

The safe-capacity problem is therefore one-sided: only the most negative
real part of \(Q^{-1}K\) can make additional capacity cross the
boundary.

\hypertarget{app-current-inference-versus-future-control}{%
\subsection*{Current inference versus future
control}\label{app:current-inference-versus-future-control}}
\addcontentsline{toc}{subsection}{Current inference versus future
control}

The scalar \(c\) in the preceding parts of Section 3 is current effective capacity and is
estimated jointly with \(Q\) and \(K\). It belongs in the current-regime
confidence region.

In this section, \(a\) is a future deployment action chosen after
observing the data. It is not an estimator of current \(c\). The
distinction is:

\[
\begin{array}{lll}
c&:&\text{uncertain current state},\\
a&:&\text{controllable future action}.
\end{array}
\tag{D.5}
\]

The joint confidence region still matters because its projection onto
\((Q,K)\) retains covariance information from the joint estimation
problem. But the future action \(a\) is not varied inside the
statistical confidence set.

\hypertarget{app-projected-upper-confidence-bound-for-intensity}{%
\subsection*{Projected upper confidence bound for
intensity}\label{app:projected-upper-confidence-bound-for-intensity}}
\addcontentsline{toc}{subsection}{Projected upper confidence bound
for intensity}

Let

\[
\mathcal C_T^{QK}
=
\left\{
(Q,K):
\exists c\text{ with }(Q,K,c)\in\mathcal C_T
\right\}
\tag{D.6}
\]

be the \((Q,K)\)-projection of the primitive region. Define the exact
upper bound

\[
U_{d,T}
=
\sup_{(Q,K)\in\mathcal C_T^{QK}}
d(Q,K).
\tag{D.7}
\]

Let the verified numerical routine return

\[
\overline U_{d,T}\ge U_{d,T}\ge0.
\tag{D.8}
\]

If \(\mathcal C_T=\varnothing\), the routine sets
\(\overline U_{d,T}=\bar k/\underline q\) and reports
primitive-estimation failure rather than authorizing unrestricted
capacity.

The same branch-and-bound architecture as Algorithm 1 applies, replacing
the margin enclosure by an enclosure of \(d\). A failed spectral
verification invokes the global bound

\[
0\le d(Q,K)
\le\frac{\bar k}{\underline q}.
\tag{D.9}
\]

\hypertarget{app-lemma-5-one-sided-coverage}{%
\subsubsection*{Lemma 5: one-sided
coverage}\label{app:lemma-5-one-sided-coverage}}
\addcontentsline{toc}{subsubsection}{Lemma 5: one-sided coverage}

Under U1--U3 and (D.8),

\[
\liminf_{T\to\infty}
\inf_{P\in\mathcal P_T^S}
P\left\{
d(Q(P),K(P))
\le\overline U_{d,T}
\right\}
\ge1-\alpha.
\tag{D.10}
\]

\hypertarget{app-proof-13}{%
\subsubsection*{Proof}\label{app:proof-13}}
\addcontentsline{toc}{subsubsection}{Proof}

On \(\vartheta(P)\in\mathcal C_T\), its \((Q,K)\) components belong to
\(\mathcal C_T^{QK}\). Therefore

\[
d(Q(P),K(P))
\le U_{d,T}
\le\overline U_{d,T}.
\]

Apply Lemma 4. \(\square\)

\hypertarget{app-planned-and-realized-capacity}{%
\subsection*{Planned and realized
capacity}\label{app:planned-and-realized-capacity}}
\addcontentsline{toc}{subsection}{Planned and realized capacity}

Let \(a_{\mathrm{des},T}\ge0\) be desired future capacity before
applying the reflexive safety constraint. It may be data-dependent.

Execution, mandate implementation or balance-sheet aggregation may cause
realized capacity \(a_{\mathrm{act},T}\) to exceed the plan. Let
\(\rho\ge0\) be a pre-specified deterministic overshoot bound:

\[
0\le a_{\mathrm{act},T}
\le(1+\rho)a_{\mathrm{plan},T}.
\tag{D.11}
\]

Choose a strict economic buffer

\[
\eta\in(0,1)
\tag{D.12}
\]

before observing the regime estimate. Define

\[
a_{\mathrm{safe},T}
=
\begin{cases}
\dfrac{1-\eta}
{2(1+\rho)\overline U_{d,T}},
&\overline U_{d,T}>0,\\[8pt]
+\infty,
&\overline U_{d,T}=0,
\end{cases}
\tag{D.13}
\]

and the planned deployment

\[
a_{\mathrm{plan},T}
=
\min\{
a_{\mathrm{des},T},
a_{\mathrm{safe},T}
\}.
\tag{D.14}
\]

The factor \(\eta\) creates distance from the mathematical boundary. The
factor \(1+\rho\) reserves room for implementation overshoot. They solve
different problems and must not be merged into a single undocumented
haircut.

\hypertarget{app-probabilistic-safety-theorem}{%
\subsection*{Probabilistic safety
theorem}\label{app:probabilistic-safety-theorem}}
\addcontentsline{toc}{subsection}{Probabilistic safety theorem}

\hypertarget{app-theorem-6a}{%
\subsubsection*{Theorem 6(a)}\label{app:theorem-6a}}
\addcontentsline{toc}{subsubsection}{Theorem 6(a)}

Suppose U1--U3, the certificate relation (D.8), and the deterministic
implementation bound (D.11) hold. Assume \(a_{\mathrm{des},T}<\infty\)
almost surely, so (D.14) and the implementation rule produce a finite
\(a_{\mathrm{act},T}\). Suppose also that the future \(Q,K\) equal the
estimands covered by \(\mathcal C_T\). Then

\[
\liminf_{T\to\infty}
\inf_{P\in\mathcal P_T^S}
P\left\{
m(Q(P),K(P),a_{\mathrm{act},T})
\ge\frac\eta2
\right\}
\ge1-\alpha.
\tag{D.15}
\]

\hypertarget{app-proof-14}{%
\subsubsection*{Proof}\label{app:proof-14}}
\addcontentsline{toc}{subsubsection}{Proof}

On the event in Lemma 5, let \(d_0=d(Q(P),K(P))\). If
\(\overline U_{d,T}>0\), then

\[
a_{\mathrm{act},T}d_0
\le
(1+\rho)a_{\mathrm{plan},T}
\overline U_{d,T}
\le
(1+\rho)a_{\mathrm{safe},T}
\overline U_{d,T}
=\frac{1-\eta}{2}.
\tag{D.16}
\]

Equation (D.4) implies

\[
m(Q(P),K(P),a_{\mathrm{act},T})
\ge
\frac12-a_{\mathrm{act},T}d_0
\ge\frac\eta2.
\]

If \(\overline U_{d,T}=0\), the covered event implies \(d_0=0\), and
(D.4) gives \(m\ge1/2\) for every finite realized deployment. Lemma 5
completes the proof. \(\square\)

\hypertarget{app-stochastic-overshoot}{%
\subsubsection*{Stochastic overshoot}\label{app:stochastic-overshoot}}
\addcontentsline{toc}{subsubsection}{Stochastic overshoot}

If (D.11) holds only on an event \(\mathcal O_T\) satisfying

\[
\liminf_T\inf_{P\in\mathcal P_T^S}
P(\mathcal O_T)\ge1-\beta,
\]

then no independence assumption is needed for the union-bound guarantee

\[
\liminf_T\inf_{P\in\mathcal P_T^S}
P\left\{
m(Q,K,a_{\mathrm{act},T})
\ge\frac\eta2
\right\}
\ge1-\alpha-\beta.
\tag{D.17}
\]

The paper reports \(\alpha\), \(\beta\), \(\eta\), and \(\rho\)
separately.

\hypertarget{app-maximum-capacity-inside-the-certified-safe-set}{%
\subsection*{Maximum capacity inside the certified safe
set}\label{app:maximum-capacity-inside-the-certified-safe-set}}
\addcontentsline{toc}{subsection}{Maximum capacity inside the
certified safe set}

Define

\[
\mathcal A_T^{\mathrm{cert}}
=
\left\{
a\in[0,a_{\mathrm{des},T}]:
(1+\rho)a\overline U_{d,T}
\le\frac{1-\eta}{2}
\right\}.
\tag{D.18}
\]

\hypertarget{app-proposition-6b}{%
\subsubsection*{Proposition 6(b)}\label{app:proposition-6b}}
\addcontentsline{toc}{subsubsection}{Proposition 6(b)}

The rule in (D.14) is the largest member of
\(\mathcal A_T^{\mathrm{cert}}\):

\[
a_{\mathrm{plan},T}
=\max\mathcal A_T^{\mathrm{cert}}.
\tag{D.19}
\]

If the ordinary, non-reflexive business value \(G_T(a)\) is
nondecreasing on \([0,a_{\mathrm{des},T}]\), then

\[
a_{\mathrm{plan},T}
\in
\arg\max_{a\in\mathcal A_T^{\mathrm{cert}}}
G_T(a).
\tag{D.20}
\]

\hypertarget{app-proof-15}{%
\subsubsection*{Proof}\label{app:proof-15}}
\addcontentsline{toc}{subsubsection}{Proof}

For \(\overline U_{d,T}>0\), (D.18) is the interval

\[
\left[
0,
\min\left\{
a_{\mathrm{des},T},
\frac{1-\eta}
{2(1+\rho)\overline U_{d,T}}
\right\}
\right].
\]

Its right endpoint is (D.14). If \(\overline U_{d,T}=0\), the constraint
is slack and the right endpoint is \(a_{\mathrm{des},T}\). Monotonicity
of \(G_T\) proves (D.20). \(\square\)

Proposition 6(b) is narrow. It shows that the rule
sacrifices no more capacity than required by the certified safety
constraint. It does not claim unrestricted welfare optimality.

\hypertarget{app-predictive-drift}{%
\subsection*{Predictive drift}\label{app:predictive-drift}}
\addcontentsline{toc}{subsection}{Predictive drift}

Theorem 6(a) is a deployment guarantee under parameter stability between
estimation and implementation. If \(Q\) or \(K\) may drift, let
\(\mathcal D_\tau\) be a predictive disturbance set for the deployment
horizon \(\tau\) and replace (D.7) by

\[
U_{d,T}^{\mathrm{pred}}
=
\sup_{\substack{(Q,K)\in\mathcal C_T^{QK}\\
(\Delta_Q,\Delta_K)\in\mathcal D_\tau}}
d(Q+\Delta_Q,K+\Delta_K).
\tag{D.21}
\]

The predictive rule is valid only if \(\mathcal D_\tau\) has its own
stated coverage guarantee and preserves \(Q+\Delta_Q\succ0\). This
extension will not be claimed in the baseline experiments unless such a
predictive set is explicitly constructed.

\hypertarget{app-asymmetric-financial-loss}{%
\subsection*{Asymmetric financial
loss}\label{app:asymmetric-financial-loss}}
\addcontentsline{toc}{subsection}{Asymmetric financial loss}

The safe-capacity theorem controls a probability, but it does not
determine how economically costly caution is. Let

\[
s(a,d)=\frac12-ad
\tag{D.22}
\]

be the destabilizing-direction slack. Let \(G(a;\zeta)\) denote ordinary
business value from deployed capacity, such as expected spread, carry or
risk premium net of conventional execution and inventory costs. The
state \(\zeta\) collects the economic calibration.

For the loss comparison, \(a_{\mathrm{des}}\) is chosen to maximize
\(G\) over the ordinary business range before imposing the
convention-boundary constraint. Hence
\(G(a_{\mathrm{des}};\zeta)-G(a;\zeta)\ge0\) on the comparison set.

Define the loss

\[
\mathcal L(a;d,\zeta)
=
\underbrace{
G(a_{\mathrm{des}};\zeta)-G(a;\zeta)
}_{\text{foregone business value}}
+
\underbrace{
F(\zeta)\mathbf 1\{s(a,d)\le0\}
+H([-s(a,d)]_+;\zeta)
}_{\text{convention exposure}}
+
\underbrace{
J(a,a_{\mathrm{prev}};\zeta)
}_{\text{adjustment cost}},
\tag{D.23}
\]

where:

\begin{itemize}
\tightlist
\item
  \(F(\zeta)\ge0\) is a fixed loss attached to crossing or touching the
  convention boundary;
\item
  \(H(0;\zeta)=0\) and \(H(\cdot;\zeta)\) is nondecreasing;
\item
  \(J\) measures turnover, funding or balance-sheet adjustment cost.
\end{itemize}

Equation (D.23) makes the asymmetry explicit:

\begin{itemize}
\tightlist
\item
  excessive deployment can incur a discrete convention-exposure loss
  plus a severity loss;
\item
  excessive caution creates foregone carry, spread or risk premium;
\item
  changing capacity itself is costly.
\end{itemize}

The baseline structural Monte Carlo uses a transparent parametric
version,

\[
G(a;\zeta)
=\pi a-\frac\gamma2a^2,
\qquad
H(x;\zeta)=\lambda x^2,
\qquad
J(a,a_{\mathrm{prev}};\zeta)
=\frac\kappa2(a-a_{\mathrm{prev}})^2,
\tag{D.24}
\]

with \(\pi,\gamma,\lambda,\kappa\ge0\). Results will be reported over a
grid of relative loss weights rather than under one preferred
calibration.

\hypertarget{app-oracle-and-regret}{%
\subsection*{Oracle and regret}\label{app:oracle-and-regret}}
\addcontentsline{toc}{subsection}{Oracle and regret}

For an admissible action set \(\mathcal A=[0,\bar a]\), define the
full-information oracle

\[
a^\star(d,\zeta)
\in
\arg\min_{a\in\mathcal A}
\mathcal L(a;d,\zeta).
\tag{D.25}
\]

For any data-dependent policy \(\delta\), its realized regret is

\[
\mathcal R(\delta;d,\zeta)
=
\mathcal L(a_\delta;d,\zeta)
-\mathcal L(a^\star(d,\zeta);d,\zeta).
\tag{D.26}
\]

The paper compares:

\begin{enumerate}
\def\labelenumi{\arabic{enumi}.}
\tightlist
\item
  desired capacity without boundary protection;
\item
  plug-in capacity based on \(\widehat d\);
\item
  pointwise-delta capacity based on a regular upper bound;
\item
  projected safe capacity from (D.14);
\item
  full withdrawal;
\item
  the infeasible oracle.
\end{enumerate}

The reported outcomes are:

\[
\begin{array}{ll}
\text{boundary violation:}&
\mathbf 1\{a_\delta d\ge1/2\},\\
\text{safety slack:}&
1/2-a_\delta d,\\
\text{deployment ratio:}&
a_\delta/a_{\mathrm{des}},\\
\text{foregone value:}&
G(a_{\mathrm{des}})-G(a_\delta),\\
\text{convention loss:}&
F\mathbf 1\{a_\delta d\ge1/2\}
+H([a_\delta d-1/2]_+),\\
\text{total regret:}&
\mathcal R(\delta;d,\zeta).
\end{array}
\tag{D.27}
\]

No regret dominance is claimed in advance. The projected rule is
designed to control boundary violation; whether that control improves
expected regret depends on the relative economic cost of convention
exposure, foregone value and adjustment.

\hypertarget{app-relation-between-classification-and-deployment}{%
\subsection*{Relation between classification and
deployment}\label{app:relation-between-classification-and-deployment}}
\addcontentsline{toc}{subsection}{Relation between classification
and deployment}

The three-way classifier answers:

\begin{quote}
On which side of the boundary is the current estimated market?
\end{quote}

The safe-capacity rule answers:

\begin{quote}
How much future capacity can be deployed while retaining a stated
probabilistic buffer?
\end{quote}

They must not be collapsed into one label. In particular:

\begin{itemize}
\tightlist
\item
  a current \(\mathrm{UNR}\) classification need not imply full
  withdrawal; (D.14) provides a conservative positive deployment when
  \(\overline U_{d,T}<\infty\);
\item
  a current \(\mathrm{SUP}_{\mathrm{spec}}\) classification indicates
  that current capacity is on the wrong side, but the action is to
  reduce future capacity below the certified cap, not to assert every
  form of economic convention equilibrium;
\item
  a current \(\mathrm{SUB}\) classification does not authorize capacity
  above \(a_{\mathrm{safe},T}\) if the proposed future deployment
  differs materially from current \(c\).
\end{itemize}

\hypertarget{app-appendix-e.-numerical-calibration-and-full-monte-carlo-evidence}{%
\section*{Appendix E. Numerical calibration and full Monte Carlo
evidence}\label{app:appendix-e.-numerical-calibration-and-full-monte-carlo-evidence}}
\addcontentsline{toc}{section}{Appendix E. Numerical calibration and
full Monte Carlo evidence}

\hypertarget{app-e.1-calibration-and-structural-design}{%
\subsection*{Calibration and structural
design}\label{app:e.1-calibration-and-structural-design}}
\addcontentsline{toc}{subsection}{Calibration and structural design}

The structural calibration study generates the observable block in
\eqref{eq:observation-model}, rather than adding noise directly to \(Q\) or \(K\). Its
stationary recursion is

\[
\begin{aligned}
Z_t&=\phi_ZZ_{t-1}+\nu_t^Z,\\
W_t&=\phi_WW_{t-1}+\nu_t^W,\\
U_t&=\phi_UU_{t-1}+\nu_t^U,
\qquad E[U_tU_t']=Q,\\
V_t&=\phi_VV_{t-1}+\nu_t^V,\\
\eta_t&=\phi_B\eta_{t-1}+\nu_t^B,\\
X_t&=M_Xg(Z_t)+\Pi W_t+V_t,\\
E_t&=L_IV_t+\xi_t,\\
Y_t&=M_Yg(Z_t)+KX_t+E_t,\\
R_t^L&=M_Qg(Z_t)+U_t,\\
B_t&=c+\eta_t.
\end{aligned}
\tag{E.1}
\]

All innovations are Gaussian. The innovations driving \(U_t,V_t\) and
\(\eta_t\) are contemporaneously correlated, while \(W_t\) is
independent of the structural error. Because \(E_t\) depends on \(V_t\),
flow is endogenous in the temporary-price equation, and \(W_t\) shifts
flow through the nonsingular matrix \(\Pi\). Thus OLS is not rescued by
construction, while the IV moment in \eqref{eq:observation-model} remains valid.

This calibration should not be used to claim that nonzero HAC
cross-blocks are empirically necessary. Under the centered Gaussian
design, the independence and parity of the IV score imply zero
population cross-covariances between the \(Q\), \(K\) and \(c\)
influence blocks, even though the underlying innovations are correlated.
The estimator nevertheless retains every sample cross-block, as required
by the general theory. E1--E4 therefore validate the joint
implementation but do not measure the cost of deleting nonzero
population cross-blocks.

\hypertarget{app-e.1.1-theory-to-design-correspondence}{%
\subsubsection*{Theory-to-design
correspondence}\label{app:e.1.1-theory-to-design-correspondence}}
\addcontentsline{toc}{subsubsection}{Theory-to-design
correspondence}

Table \ref{tab:app-design-correspondence} maps the numerical design to the sufficient class in Section 3.1.

\begin{longtable}[]{@{}
  >{\raggedright\arraybackslash}p{(\columnwidth - 4\tabcolsep) * \real{0.333}}
  >{\raggedright\arraybackslash}p{(\columnwidth - 4\tabcolsep) * \real{0.333}}
  >{\raggedright\arraybackslash}p{(\columnwidth - 4\tabcolsep) * \real{0.333}}@{}}
\caption{Assumption-to-design correspondence for the full simulation.}\label{tab:app-design-correspondence}\\
\endfirsthead
\toprule\noalign{}
\begin{minipage}[b]{\linewidth}\raggedright
Condition
\end{minipage} & \begin{minipage}[b]{\linewidth}\raggedright
Pre-specified design
\end{minipage} & \begin{minipage}[b]{\linewidth}\raggedright
Status and diagnostic
\end{minipage} \\
\midrule\noalign{}
\endhead
\bottomrule\noalign{}
\endlastfoot
A1: dependence & Stable Gaussian Markov recursion with
\(\max|\phi|=0.550\) & Geometrically mixing. Each series starts at zero and
discards 250 observations; the largest marginal covariance discrepancy
is bounded by \(0.550^{500}=1.520\times10^{-130}\). This is covered by the
geometrically coupled initialization extension of A1. \\
A2: moments and design & \(p=2\), \(q=2\), fixed dictionary
\(g(Z)=(1,Z)'\), Gaussian innovations & All required moments exist and
dimensions do not grow. \\
A3: structural moments & \(U_t\) is independent of \(Z_t\); \(W_t\) is
independent of \(E_t\); \(E_t=L_IV_t+\xi_t\) & Conditional-risk and IV
moments hold for the stationary target; endogeneity remains because
\(V_t\) enters both \(X_t\) and \(E_t\). \\
A4: strong IV & \(M_{XW}=\Pi\Omega_W\) & Population minimum singular
value \(0.576\); across full experiment, sample minimum \(0.303\)
and mean \(0.565\). No weak-IV sequence is simulated. \\
A5: interiority & \(\lambda(Q)=(0.795,1.305)\), \(c=1\),
\(\max\|K\|_F=2.189\) & Strictly inside the pre-specified bounds
\(q\in[0.200,3]\), \(c\in[0.400,1.600]\), \(\|K\|_F\le5\). \\
A6: HAC & Bartlett kernel and \(b_T=\lfloor4(T/100)^{2/9}\rfloor\) &
\(b_T=4\) at \(T=250\) and \(6\) at \(T=1000\). All 60,000 HAC matrices
were positive definite and the declared \(10^{-10}\) eigenvalue
safeguard was never activated. \\
\end{longtable}

The stable Gaussian construction and positive idiosyncratic variance in
each score block make the population long-run covariance finite and
nonsingular. The finite-sample diagnostics corroborate, but do not prove
beyond this DGP, the uniform lower bound in A6. The local and Jordan
sequences alter only \(K\); all dependence, moment, identification,
covariance and compactness constants remain common across the triangular
design.

The pre-specified calibration study includes six geometries: symmetric
separated, nonnormal diagonalizable, complex active pair, semisimple
tie, near-Jordan transition and exact Jordan. For each
\(T\in\{250,1000\}\), it uses

\[
m_T\in
\left\{
-C,-\frac{0.5}{\sqrt T},0,
\frac{0.5}{\sqrt T},C
\right\},
\qquad C=0.75.
\tag{E.2}
\]

An initial calibration used \(C=0.15\), but those nominal fixed-distance
cells remained inside the joint eight-parameter Wald uncertainty region.
Before the reported calibration study, one common \(C=0.75\) was fixed for
all geometries, signs and sample sizes; the local and zero sequences were
unchanged.

The economic layer fixes \(G(a)=a-a^2/2\), \(a_{\mathrm{des}}=1\),
\(a_{\mathrm{prev}}=0.75\), \(\eta=0.10\) and \(\rho=0.05\). It compares
low, baseline and high convention-loss calibrations without claiming
regret dominance.

\hypertarget{app-e.4-full-e1e4-monte-carlo-experiment}{%
\subsection*{Full E1--E4 Monte Carlo
experiment}\label{app:e.4-full-e1e4-monte-carlo-experiment}}
\addcontentsline{toc}{subsection}{Full E1--E4 Monte Carlo
experiment}

The full experiment design fixes 1,000 replications in each of the 60
cells, so the worst-case cell-level Monte Carlo standard error for a binary
statistic is at most \(0.5/\sqrt{1000}=0.016\). Each replication is identified
by sample size, geometry, distance, and replication number. Both the
primitive-region and spectral certificates are checked before a replication
enters the aggregates.

Because the grid is fixed and equally weighted, the global estimand is
the average of the 60 cell means. For any replication statistic \(H\),
its Monte Carlo standard error is therefore computed as

\[
\widehat{\operatorname{se}}_{\mathrm{MC}}(\bar H)
=
\left\{
\frac1{J^2}\sum_{j=1}^J\frac{s_j^2}{n_j}
\right\}^{1/2},
\qquad J=60,
\quad n_j=1000,
\tag{E.3}
\]

where \(s_j^2\) is the within-cell sample variance. A standard deviation
computed after pooling all raw rows would mix cross-cell design
dispersion with simulation noise and is not used below.

Equation (E.3) treats distinct design keys as independent. The
pre-specified run contains two disclosed collisions in its 32-bit seed
map, both across rather than within design cells. For a statistic
bounded in \([0,1]\), the largest possible additional standard-error
component from two perfectly correlated pairs is \(1/60000=0.002\)
percentage points. The collisions are therefore retained and disclosed
rather than removed by selective rerunning; future experiments require a
collision-free seed map.

Table \ref{tab:app-global-mc} reports the global results. The full experiment
contains 60,000 successful inference replications, with no failed primitive
estimates and no rejected spectral certificates. The corresponding financial
comparison contains 1,260,000 evaluations, and the reported summaries use the
cell-weighted Monte Carlo standard errors in (E.3).

\begin{longtable}[]{@{}
  >{\raggedright\arraybackslash}p{(\columnwidth - 12\tabcolsep) * \real{0.111}}
  >{\raggedleft\arraybackslash}p{(\columnwidth - 12\tabcolsep) * \real{0.148}}
  >{\raggedleft\arraybackslash}p{(\columnwidth - 12\tabcolsep) * \real{0.148}}
  >{\raggedleft\arraybackslash}p{(\columnwidth - 12\tabcolsep) * \real{0.148}}
  >{\raggedleft\arraybackslash}p{(\columnwidth - 12\tabcolsep) * \real{0.148}}
  >{\raggedleft\arraybackslash}p{(\columnwidth - 12\tabcolsep) * \real{0.148}}
  >{\raggedleft\arraybackslash}p{(\columnwidth - 12\tabcolsep) * \real{0.148}}@{}}
\caption{Full global Monte Carlo comparison.}\label{tab:app-global-mc}\\
\endfirsthead
\toprule\noalign{}
\begin{minipage}[b]{\linewidth}\raggedright
Method
\end{minipage} & \begin{minipage}[b]{\linewidth}\raggedleft
Coverage
\end{minipage} & \begin{minipage}[b]{\linewidth}\raggedleft
MC SE
\end{minipage} & \begin{minipage}[b]{\linewidth}\raggedleft
Wrong resolved
\end{minipage} & \begin{minipage}[b]{\linewidth}\raggedleft
MC SE
\end{minipage} & \begin{minipage}[b]{\linewidth}\raggedleft
Unresolved
\end{minipage} & \begin{minipage}[b]{\linewidth}\raggedleft
MC SE
\end{minipage} \\
\midrule\noalign{}
\endhead
\bottomrule\noalign{}
\endlastfoot
Pointwise delta & 86.510\% & 0.133 pp & 4.523\% & 0.077 pp &
51.900\% & 0.111 pp \\
Deterministic norm & 100.000\% & 0.000 pp & 0.000\% & 0.000 pp &
91.948\% & 0.040 pp \\
Projected certified envelope & 99.992\% & 0.004 pp & 0.003\% & 0.002
pp & 60.170\% & 0.018 pp \\
\end{longtable}

The plug-in classifier is wrong in 34.863\% of the full design. The
projected procedure therefore occupies the intended middle ground.
Relative to the deterministic norm bound, it resolves far more cells
while retaining near-complete empirical coverage. Relative to pointwise
delta inference, it materially increases coverage and reduces wrong
resolved declarations from 4.523\% to 0.003\%.

Table \ref{tab:app-distance-results} reports the distance-specific results. The two projected wrong resolved declarations both occur at the exact
boundary and coincide with projected-envelope coverage misses. There are
five projected coverage misses in total. Counts are reported because
rounding the wrong resolved rate to zero would conceal these cases.

\begin{longtable}[]{@{}
  >{\raggedright\arraybackslash}p{(\columnwidth - 10\tabcolsep) * \real{0.130}}
  >{\raggedleft\arraybackslash}p{(\columnwidth - 10\tabcolsep) * \real{0.174}}
  >{\raggedleft\arraybackslash}p{(\columnwidth - 10\tabcolsep) * \real{0.174}}
  >{\raggedleft\arraybackslash}p{(\columnwidth - 10\tabcolsep) * \real{0.174}}
  >{\raggedleft\arraybackslash}p{(\columnwidth - 10\tabcolsep) * \real{0.174}}
  >{\raggedleft\arraybackslash}p{(\columnwidth - 10\tabcolsep) * \real{0.174}}@{}}
\caption{Full results by distance from the boundary.}\label{tab:app-distance-results}\\
\endfirsthead
\toprule\noalign{}
\begin{minipage}[b]{\linewidth}\raggedright
Distance
\end{minipage} & \begin{minipage}[b]{\linewidth}\raggedleft
Projected coverage
\end{minipage} & \begin{minipage}[b]{\linewidth}\raggedleft
Wrong resolved
\end{minipage} & \begin{minipage}[b]{\linewidth}\raggedleft
Unresolved
\end{minipage} & \begin{minipage}[b]{\linewidth}\raggedleft
Classification complete
\end{minipage} & \begin{minipage}[b]{\linewidth}\raggedleft
Capacity precise
\end{minipage} \\
\midrule\noalign{}
\endhead
\bottomrule\noalign{}
\endlastfoot
Fixed supercritical & 99.992\% & 0.000\% & 0.000\% & 100.000\% &
100.000\% \\
Local supercritical & 99.992\% & 0.000\% & 99.933\% & 99.942\% &
99.900\% \\
Boundary & 99.983\% & 0.017\% & 99.983\% & 100.000\% & 99.767\% \\
Local subcritical & 99.992\% & 0.000\% & 99.975\% & 100.000\% &
99.733\% \\
Fixed subcritical & 100.000\% & 0.000\% & 0.958\% & 99.817\% &
99.933\% \\
\end{longtable}

The local and exact-boundary nonresolution rates are statistical
abstention, not failed forced classification. Fixed-distance cells
resolve almost completely. This ordering is consistent with the uniform
boundary theory; the simulation illustrates rather than proves the
asymptotic resolution statements.

Table \ref{tab:app-geometry-results} reorganizes the same pre-specified output by spectral geometry and sample size. Each main percentage averages equally over the
five distance cells, with 1,000 replications per cell. The last column
isolates the two fixed-distance cells so that difficulty caused by
geometry is not hidden by the three local or boundary cells, which are
designed to abstain.

\begin{longtable}[]{@{}
  >{\raggedleft\arraybackslash}p{(\columnwidth - 10\tabcolsep) * \real{0.174}}
  >{\raggedright\arraybackslash}p{(\columnwidth - 10\tabcolsep) * \real{0.130}}
  >{\raggedleft\arraybackslash}p{(\columnwidth - 10\tabcolsep) * \real{0.174}}
  >{\raggedleft\arraybackslash}p{(\columnwidth - 10\tabcolsep) * \real{0.174}}
  >{\raggedleft\arraybackslash}p{(\columnwidth - 10\tabcolsep) * \real{0.174}}
  >{\raggedleft\arraybackslash}p{(\columnwidth - 10\tabcolsep) * \real{0.174}}@{}}
\caption{Full results by geometry and sample size.}\label{tab:app-geometry-results}\\
\endfirsthead
\toprule\noalign{}
\begin{minipage}[b]{\linewidth}\raggedleft
\(T\)
\end{minipage} & \begin{minipage}[b]{\linewidth}\raggedright
Spectral geometry
\end{minipage} & \begin{minipage}[b]{\linewidth}\raggedleft
Coverage
\end{minipage} & \begin{minipage}[b]{\linewidth}\raggedleft
Wrong resolved
\end{minipage} & \begin{minipage}[b]{\linewidth}\raggedleft
Unresolved, all distances
\end{minipage} & \begin{minipage}[b]{\linewidth}\raggedleft
Unresolved, fixed distances
\end{minipage} \\
\midrule\noalign{}
\endhead
\bottomrule\noalign{}
\endlastfoot
250 & Symmetric separated & 100.000\% & 0.000\% & 60.000\% &
0.000\% \\
250 & Nonnormal diagonalizable & 100.000\% & 0.000\% & 60.240\% &
0.600\% \\
250 & Complex active pair & 99.980\% & 0.000\% & 59.980\% &
0.000\% \\
250 & Semisimple tie & 100.000\% & 0.000\% & 59.940\% & 0.000\% \\
250 & Near-Jordan transition & 99.980\% & 0.000\% & 60.320\% &
0.850\% \\
250 & Exact Jordan & 100.000\% & 0.000\% & 61.700\% & 4.300\% \\
1000 & Symmetric separated & 99.980\% & 0.020\% & 59.940\% &
0.000\% \\
1000 & Nonnormal diagonalizable & 100.000\% & 0.000\% & 60.000\% &
0.000\% \\
1000 & Complex active pair & 99.980\% & 0.000\% & 59.980\% &
0.000\% \\
1000 & Semisimple tie & 100.000\% & 0.000\% & 59.980\% & 0.000\% \\
1000 & Near-Jordan transition & 100.000\% & 0.000\% & 60.000\% &
0.000\% \\
1000 & Exact Jordan & 99.980\% & 0.020\% & 59.960\% & 0.000\% \\
\end{longtable}

At \(T=250\), the fixed-distance unresolved rate rises from zero in the
symmetric, complex and semisimple designs to 0.60\% under nonnormality,
0.85\% near the Jordan transition and 4.30\% at the exact Jordan block.
At \(T=1000\), every fixed-distance geometry resolves. The two wrong
resolved declarations are the already disclosed exact-boundary misses in
the symmetric and exact-Jordan rows; they are not fixed-distance
failures.

The solver terminates adaptively in 59,892 runs and exhausts the
pre-specified 400-box budget in 108 runs. Classification completes in
99.952\% of replications and the capacity-precision target is met in
99.867\%. Twenty-nine runs do not complete classification, 80 do not
attain the capacity-action tolerance, and one misses both components.
The mean capacity-action gap is 0.003, while the maximum is
0.14. Budget-exhausted outputs remain outward-certified but are
not described as having met the adaptive precision selection criterion.

The exact two-dimensional Routh--Hurwitz ball certificate is used in
58,895 replications; the conservative symmetric-part fallback is used in
1,105. There are no primitive-estimation failures and no retained
fallback-active boxes. The minimum certified guaranteed margin is
0.05, equal to the target 0.05 up to outward numerical
rounding. Two 32-bit seed collisions occur among the 60,000 design keys.
They are disclosed in Appendix F; the pre-specified outcomes are not
selectively rerun.

Table \ref{tab:app-capacity-results} gives the financial comparison under the baseline loss convention.

\begin{longtable}[]{@{}
  >{\raggedright\arraybackslash}p{(\columnwidth - 8\tabcolsep) * \real{0.158}}
  >{\raggedleft\arraybackslash}p{(\columnwidth - 8\tabcolsep) * \real{0.2105}}
  >{\raggedleft\arraybackslash}p{(\columnwidth - 8\tabcolsep) * \real{0.2105}}
  >{\raggedleft\arraybackslash}p{(\columnwidth - 8\tabcolsep) * \real{0.2105}}
  >{\raggedleft\arraybackslash}p{(\columnwidth - 8\tabcolsep) * \real{0.2105}}@{}}
\caption{Full capacity and regret comparison.}\label{tab:app-capacity-results}\\
\endfirsthead
\toprule\noalign{}
\begin{minipage}[b]{\linewidth}\raggedright
Policy
\end{minipage} & \begin{minipage}[b]{\linewidth}\raggedleft
Planned capacity
\end{minipage} & \begin{minipage}[b]{\linewidth}\raggedleft
Violation rate
\end{minipage} & \begin{minipage}[b]{\linewidth}\raggedleft
Mean regret
\end{minipage} & \begin{minipage}[b]{\linewidth}\raggedleft
Regret MC SE
\end{minipage} \\
\midrule\noalign{}
\endhead
\bottomrule\noalign{}
\endlastfoot
Oracle & 0.857 & 0.000\% & 0.000 & 0.000 \\
Projected safe & 0.541 & 0.000\% & 0.104 & 0.000 \\
Deterministic norm & 0.342 & 0.000\% & 0.196 & 0.000 \\
Full withdrawal & 0.000 & 0.000\% & 0.489 & 0.000 \\
Pointwise delta & 0.666 & 1.758\% & 0.066 & 0.000 \\
Plug-in & 0.758 & 6.233\% & 0.083 & 0.001 \\
Desired unprotected & 1.000 & 58.333\% & 1.673 & 0.000 \\
\end{longtable}

The projected-safe policy records zero boundary violations in every
design cell and improves deployment and regret relative to the other
non-oracle zero-violation policies. It does not dominate policies that
accept positive violation risk under the baseline or low convention-loss
calibration. Under the high convention-loss calibration, projected-safe
mean regret is 0.107, below pointwise delta at 0.12 and plug-in
at 0.271. The evidence is therefore a safety-performance tradeoff,
not unconditional regret dominance.

The completed experiment supports finite-sample claims only for the
pre-specified simulated design. It does not establish universal
coverage, zero wrong-classification probability, universal adaptive
completion, real-market capacity validity, high-dimensional
certification, unconditional regret dominance or causal identification
from observed market data.

\hypertarget{app-e.5-empirical-q-numerical-stress-test}{%
\subsection*{\texorpdfstring{Empirical-\(Q\)}{Empirical-Q} numerical stress
test}\label{app:e.5-empirical-q-numerical-stress-test}}
\addcontentsline{toc}{subsection}{Empirical-Q numerical stress test}

This robustness exercise replaces the simulated risk matrix and its
uncertainty by rolling estimates from observed return panels while
leaving the structural cross-impact operator and capacity calibration
declared rather than estimated. It is therefore a numerical-realism
stress test of the certified projection procedure, not a market
application of the structural model.

The inputs are daily prices for 150 equities and 1,336 Spanish bonds.
Assets require at least 80\% return coverage. Within each universe, three
pairs are selected at the 10th, 50th and 90th percentiles of absolute
training-sample correlation. Ten rolling windows per pair produce 60
two-dimensional \(Q\) cases. Equity windows contain 252 returns and bond
windows 126. For each window, uncertainty is the Bartlett--HAC covariance
of \(\operatorname{vech}(Q)\), computed from centered return outer
products.

The pre-specified E1--E4 margin design is then applied without retuning. The
three declared \(K\) geometries are symmetric separated, nonnormal
diagonalizable and exact Jordan; \(c=1\). Target margins are
\(-0.75\), \(-0.5/\sqrt n\), \(0\), \(0.5/\sqrt n\) and \(0.75\).
The 95\% primitive radius, 400-box budget, adaptive stopping rule and
ball-aware relaxation are unchanged. This yields \(60\times3\times5=900\)
certified solver calls. All 900 certificates verify, with no primitive
estimation failures.

Table \ref{tab:empirical-q} summarizes the empirical-risk stress exercise.

\begin{longtable}[]{@{}
  >{\raggedright\arraybackslash}p{0.300\columnwidth}
  >{\raggedleft\arraybackslash}p{0.210\columnwidth}
  >{\raggedleft\arraybackslash}p{0.210\columnwidth}
  >{\raggedleft\arraybackslash}p{0.210\columnwidth}@{}}
\caption{Resolution of the empirical-risk stress exercise by universe.}\label{tab:empirical-q}\\
\endfirsthead
\toprule\noalign{}
Universe & Fixed-distance correct & Budget exhausted & Wrong resolved \\
\midrule\noalign{}
\endhead
\bottomrule\noalign{}
\endlastfoot
Equities & 95.000\% & 5.000\% & 0.000\% \\
Spanish bonds & 38.890\% & 30.000\% & 0.000\% \\
All cases & 66.940\% & 13.890\% & 0.000\% \\
\end{longtable}

Every local and exact-boundary case remains unresolved, as required by
the conservative three-way decision rule in this stress design. Across
all margin cells, the certified procedure resolves 26.78\%, the
pointwise delta comparator resolves 21.22\%, and delta resolves while
the certified projection abstains in 3.89\%. The experiment therefore
does not obtain apparent resolution by accepting wrong fixed-distance
signs.

The lower bond resolution is consistent with, but does not by itself
identify, the combined effect of shorter windows, stale-price zeros and
larger HAC uncertainty. More importantly, observed prices identify only
the risk geometry used here. They do not identify \(K\), \(c\), signed
flow, an excluded shifter, the resilience horizon or a balance-sheet
capacity conversion. Accordingly, this subsection supports robustness
of the numerical certification under observed \(Q\) geometry; it is not
evidence that a real market has been classified relative to the
reflexive stability boundary.

\begin{center}\rule{0.5\linewidth}{0.5pt}\end{center}

\hypertarget{app-appendix-f.-data-and-replication-protocols}{%
\section*{Appendix F. Data and replication protocols
}\label{app:appendix-f.-data-and-replication-protocols}}
\addcontentsline{toc}{section}{Appendix F. Data and replication protocols}

\hypertarget{app-provenance}{%
\subsection*{Structural Monte Carlo}\label{app:provenance}}
\addcontentsline{toc}{subsection}{Structural Monte Carlo}

The simulation crosses two sample sizes, six spectral geometries, and five
distances from the boundary, giving 60 equally weighted design cells. Each cell
contains 1,000 replications. To replicate the experiment, first generate the
Gaussian Markov system in Appendix E, discard 250 observations, and retain the
next \(T\in\{250,1000\}\). Estimate \(Q\), \(K\), and \(c\) jointly; construct
the Bartlett--HAC covariance with the bandwidth in Table
\ref{tab:app-design-correspondence}; and form the 95\% primitive Wald region.
Apply Algorithm 1 to obtain the certified margin interval, the three-way regime
classification, and the intensity bound. Finally, evaluate each of the 21
declared loss-policy combinations using the same primitive draw and aggregate
within cells before assigning equal weight across the 60 cells. Monte Carlo
standard errors use (E.3), not the variance of the pooled heterogeneous rows.

The six geometries are symmetric separated, nonnormal diagonalizable, complex
active pair, semisimple tie, near-Jordan transition, and exact Jordan. The five
distances are \(-0.75\), \(-0.5/\sqrt T\), zero, \(0.5/\sqrt T\), and
\(0.75\). The financial layer fixes \(G(a)=a-a^2/2\), desired capacity 1,
previous capacity 0.75, statistical buffer 0.10, and implementation buffer
0.05. No design cell is rerun or reweighted after inspection.

\subsection*{Observed-risk data and sampling}
\addcontentsline{toc}{subsection}{Observed-risk data and sampling}

The observed-risk stress uses two daily closing-price panels. The equity panel
contains 150 securities from 3 May 2010 through 23 February 2024; the Spanish
bond panel contains 1,336 securities from 29 May 2014 through 20 December 2018.
Assets must have at least 80\% return coverage. Within each universe, the pairs
closest to the 10th, 50th, and 90th percentiles of absolute training-sample
correlation are selected. Ten nonoverlapping rolling windows per pair produce
60 risk-matrix cases: equity windows contain 252 returns and bond windows 126.
For every case, \(Q\) is the centered return covariance and its uncertainty is
the Bartlett--HAC covariance of \(\operatorname{vech}(Q)\).

The risk matrices are combined with symmetric separated, nonnormal
diagonalizable, and exact-Jordan impact geometries, five target margins, and
\(c=1\), producing 900 certified calculations. Cross-impact and capacity are
imposed rather than estimated, so this exercise measures sensitivity to
observed risk geometry and information quality; it is not a classification of
either market.

\hypertarget{app-independent-integrity-checks}{%
\subsection*{Completion checks
}\label{app:independent-integrity-checks}}
\addcontentsline{toc}{subsection}{Completion checks}

The following checks are applied before aggregation.

\begin{itemize}
\tightlist
\item
  Every cell contains replication indices 0 through 999 exactly once.
\item
  The design contains exactly 60,000 unique combinations of sample size,
  geometry, distance, and replication.
\item
  Every retained observation has a valid primitive region, an outward spectral
  enclosure, a classifier, and a capacity bound.
\item
  The 60 inference groups and 1,260 financial groups are calculated directly
  from the replication-level outcomes using (E.3).
\item
  Every primitive draw is evaluated under the same 21 loss-policy
  combinations.
\item
  There are no primitive-estimation failures and no retained
  fallback-active boxes.
\item
  The exact two-dimensional Routh--Hurwitz certificate is used in 58,895
  cases; the conservative symmetric-part fallback is used in 1,105 cases.
\item
  The minimum certified guaranteed margin is 0.05, equal to the target 0.05 up
  to outward numerical rounding.
\end{itemize}

Two 32-bit seed collisions occur among 60,000 design keys. This is
consistent with the birthday-collision scale for a 32-bit seed space and is
too sparse to change the numerical conclusions. A new replication should use
a collision-free 64-bit seed mapping.

\hypertarget{app-f.1-seed-collision-disclosure}{%
\subsection*{Seed-collision
disclosure}\label{app:f.1-seed-collision-disclosure}}
\addcontentsline{toc}{subsection}{Seed-collision disclosure}

The 32-bit seed map generated exactly two collisions among the 60,000
design keys:

\begin{itemize}
\tightlist
\item
  seed \(42{,}423{,}108\):
  \((T=250,\ \text{nonnormal diagonalizable},\ \text{local super},\ r=950)\)
  and
  \((T=1000,\ \text{nonnormal diagonalizable},\ \text{local sub},\ r=784)\);
\item
  seed \(1{,}909{,}885{,}036\):
  \((T=250,\ \text{nonnormal diagonalizable},\ \text{fixed sub},\ r=356)\)
  and \((T=1000,\ \text{semisimple tie},\ \text{fixed sub},\ r=26)\).
\end{itemize}

The collisions are disclosed rather than selectively rerun. They do not
create duplicate design keys, missing cells, rejected certificates, or
altered aggregate weights.

\hypertarget{app-five-projected-coverage-misses}{%
\subsection*{Five projected coverage
misses}\label{app:five-projected-coverage-misses}}
\addcontentsline{toc}{subsection}{Five projected coverage misses}

The misses occur in five distinct cells:

\begin{enumerate}
\def\labelenumi{\arabic{enumi}.}
\tightlist
\item
  \(T=250\), complex pair, local supercritical, replication 38.
\item
  \(T=1000\), symmetric separated, boundary, replication 93.
\item
  \(T=250\), near-Jordan transition, local subcritical, replication
  664.
\item
  \(T=1000\), complex pair, fixed supercritical, replication 728.
\item
  \(T=1000\), exact Jordan, boundary, replication 739.
\end{enumerate}

The boundary misses at replications 93 and 739 are the only two projected wrong
resolved classifications. The empirical projected coverage remains 99.992\%,
far above the nominal confidence level; counts are retained because rounding
the percentages would conceal these events.

\hypertarget{references}{%
\section*{References}\label{references}}
\addcontentsline{toc}{section}{References}
\small

Andrews, D. W. K. (1991). ``Heteroskedasticity and Autocorrelation
Consistent Covariance Matrix Estimation.'' \emph{Econometrica} 59(3),
817--858.
\href{https://EconPapers.repec.org/RePEc:ecm:emetrp:v:59:y:1991:i:3:p:817-58}{Bibliographic
record}.

Andrews, D. W. K., and X. Cheng (2012). ``Estimation and Inference With
Weak, Semi-Strong, and Strong Identification.'' \emph{Econometrica}
80(5), 2153--2211.
\href{https://cowles.yale.edu/sites/default/files/2022-08/d1773-r.pdf}{Open
Cowles version}.

Benzaquen, M., I. Mastromatteo, Z. Eisler, and J.-P. Bouchaud (2017).
``Dissecting Cross-Impact on Stock Markets: An Empirical Analysis.''
\emph{Journal of Statistical Mechanics: Theory and Experiment}, 023406.
\href{https://doi.org/10.1088/1742-5468/aa53f7}{DOI}.

Bentkus, V. (1986). ``Dependence of the Berry--Esseen Estimate on the
Dimension.'' \emph{Lithuanian Mathematical Journal} 26, 110--114.
\href{https://doi.org/10.1007/BF00966143}{DOI}.

Burke, J. V., and M. L. Overton (2001). ``Variational Analysis of
Non-Lipschitz Spectral Functions.'' \emph{Mathematical Programming} 90,
317--351.
\href{https://doi.org/10.1007/s102080010008}{DOI}.

Chernozhukov, V., H. Hong, and E. Tamer (2007). ``Estimation and
Confidence Regions for Parameter Sets in Econometric Models.''
\emph{Econometrica} 75(5), 1243--1284.
\href{https://dspace.mit.edu/bitstream/handle/1721.1/63545/estimationconfid00cher.pdf}{Open
MIT manuscript}.

de Jong, R. M., and J. Davidson (2000). ``Consistency of Kernel
Estimators of Heteroscedastic and Autocorrelated Covariance Matrices.''
\emph{Econometrica} 68(2), 407--424.
\href{https://ideas.repec.org/a/ecm/emetrp/v68y2000i2p407-424.html}{Bibliographic
record}.

Doukhan, P. (1994). \emph{Mixing: Properties and Examples}. Lecture
Notes in Statistics 85. Springer, New York.
\href{https://link.springer.com/book/10.1007/978-1-4612-2642-0}{Springer
record}.

Dufour, J.-M. (1997). ``Some Impossibility Theorems in Econometrics with
Applications to Structural and Dynamic Models.'' \emph{Econometrica}
65(6), 1365--1388.
\href{https://EconPapers.repec.org/RePEc:ecm:emetrp:v:65:y:1997:i:6:p:1365-1388}{Bibliographic
record}.

Fang, Z., and A. Santos (2019). ``Inference on Directionally
Differentiable Functions.'' \emph{Review of Economic Studies} 86(1),
377--412. \href{https://arxiv.org/abs/1404.3763}{Open manuscript}.

Gatheral, J. (2010). ``No-Dynamic-Arbitrage and Market Impact.''
\emph{Quantitative Finance} 10(7), 749--759.
\href{https://doi.org/10.1080/14697680903373692}{DOI}.

Gebbie, T. (2026). ``Reflexivity from Hierarchical Causality.'' Working
paper, Department of Statistical Sciences, University of Cape Town.

Glasserman, P., and X. Xu (2014). ``Robust Risk Measurement and Model
Risk.'' \emph{Quantitative Finance} 14(1), 29--58.
\href{https://doi.org/10.1080/14697688.2013.822989}{DOI}.

Imbens, G. W., and C. F. Manski (2004). ``Confidence Intervals for
Partially Identified Parameters.'' \emph{Econometrica} 72(6),
1845--1857.
\href{https://www.cemmap.ac.uk/wp-content/uploads/2020/08/CWP0903.pdf}{Open
working paper}.

Hinrichsen, D., and A. J. Pritchard (1986). ``Stability Radii of Linear
Systems.'' \emph{Systems \& Control Letters} 7(1), 1--10.
\href{https://wrap.warwick.ac.uk/id/eprint/25716/}{Bibliographic record}.

Kato, T. (1995). \emph{Perturbation Theory for Linear Operators}.
Classics in Mathematics. Springer, Berlin.
\href{https://link.springer.com/book/10.1007/978-3-642-66282-9}{Springer
record}.

Mastromatteo, I., M. Benzaquen, Z. Eisler, and J.-P. Bouchaud (2017).
``Trading Lightly: Cross-Impact and Optimal Portfolio Execution.''
\href{https://arxiv.org/abs/1702.03838}{arXiv:1702.03838}.

Mikusheva, A. (2007). ``Uniform Inference in Autoregressive Models.''
\emph{Econometrica} 75(5), 1411--1452.
\href{https://economics.mit.edu/sites/default/files/publications/uniform_inferences.pdf}{Open
manuscript}.

Moore, R. E., R. B. Kearfott, and M. J. Cloud (2009). \emph{Introduction
to Interval Analysis}. SIAM, Philadelphia.
\href{https://epubs.siam.org/doi/book/10.1137/1.9780898717716}{SIAM
record}.

Rendl, F., and H. Wolkowicz (1997). ``A Semidefinite Framework for Trust
Region Subproblems with Applications to Large Scale Minimization.''
\emph{Mathematical Programming} 77, 273--299.
\href{https://doi.org/10.1007/BF02614438}{DOI}.

Rodríguez Domínguez, A. (2026a). ``Dynamic Causal Portfolio Choice:
Hedging the Rotation of the Common-Driver Manifold.'' arXiv:2607.06702
{[}q-fin.PM{]}.
\href{https://arxiv.org/abs/2607.06702}{Public preprint}.

Rodríguez Domínguez, A. (2026b). ``The Market's Conditioning
Representation: Equilibrium, Crowding, and Convention Multiplicity.''
arXiv:2608.18299 {[}q-fin.PM{]}, Lemma 1 and Theorem 5.
\href{https://arxiv.org/abs/2608.18299}{Public preprint}.

Rodríguez Domínguez, A. (2026c). ``Switching Frictions, Heterogeneous
Trading Horizons, and Long-Memory Order Flow.'' Public preprint.

Rump, S. M. (2022). ``Verified Error Bounds for All Eigenvalues and
Eigenvectors of a Matrix.'' Unpublished manuscript.
\href{https://www.tuhh.de/ti3/paper/rump/Ru22a.pdf}{Open manuscript}.

Schneider, M., and F. Lillo (2019). ``Cross-Impact and
No-Dynamic-Arbitrage.'' \emph{Quantitative Finance} 19(1), 137--154.
\href{https://doi.org/10.1080/14697688.2018.1467033}{DOI}.

Tomas, M., I. Mastromatteo, and M. Benzaquen (2022). ``How to Build a
Cross-Impact Model from First Principles: Theoretical Requirements and
Empirical Results.''
\href{https://arxiv.org/abs/2004.01624}{arXiv:2004.01624}.

\normalsize
\hypertarget{declarations}{%
\subsection*{Declarations}\label{declarations}}
\addcontentsline{toc}{subsection}{Declarations}

\textbf{Funding:} The author received no specific funding for this
research.

\textbf{Competing interests:} The author is employed by a financial
institution that trades asset classes related to the subject of this
study. The analysis uses no client data, and the views expressed are the
author's own.

\textbf{Data availability:} The observed-risk exercise uses the equity and
Spanish-bond price panels described in Appendix F. No client or other
proprietary data are used. The experiment-specific transformations and sampling
rules are stated in the appendices.

\end{document}